\documentclass[reprint,twocolumn,superscriptaddress,nofootinbib,amsmath,amssymb,aps,prl]{revtex4-1}%showpacs,tightenlines
\usepackage{amsmath}
\usepackage{amssymb}
\usepackage{changepage}
\usepackage{epsfig}
\usepackage{epstopdf}
\usepackage{bm}
\usepackage[colorlinks,citecolor=blue,linkcolor=blue,urlcolor=blue,]{hyperref}
\usepackage{lipsum}

\graphicspath{{figure0/}}

\begin{document}
	
	\title{Purcell effect and quantum Zeno effect suppressed self-discharging of quantum battery}
	\author{Da-Wei Liu, Guo-Qing Tian, Zi-Hao Li, Shi-Qi Gan, Ying Wu, and Liu-Gang Si}
	
	\email{siliugang@hust.edu.cn}
	\affiliation{School of Physics, Huazhong University of Science and Technology, Wuhan 430074, People’s Republic of China}
	\begin{abstract}
		Quantum batteries (QB), as an energy storage and transfer device, not only show obvious advantages compared to classical electrochemical batteries, but also have important applications in quantum information. %This purcell effect can significantly affect the spontaneous emission rate of atoms by changing the spectral density. 
		Self-discharging is a central obstacle to storing useful work in open QB, especially when the charger itself provides an unavoidable loss channel. Here we show that such charger-induced loss can be converted into a protection mechanism by combining Purcell effect with quantum Zeno effect.
%		Here, we propose a virtual photon charging protocol that forms an effective second-order coupling between the near-resonant QB and the charger through a large detuned auxiliary cavity, thereby transfering energy from the charger to the QB. 
		We reveal that the virtual photon process and the Purcell effect can induce the strong coupling regime to the quantum Zeno regime, in which the stronger the dissipation of the charger, the weaker the self-discharging effect of the QB. 
		%The virtual photon process and the purcell effect will induce the quantum Zeno phenomenon, where the originally strongly coupled system is mapped to the quantum Zeno regime by the detuning.
		As a result, the dissipation caused by the charger to the QB can be suppressed four orders of magnitude in our scheme. Meanwhile, the quantum Zeno effect induced by the Purcell effect can also avoid the energy backflow between the QB and the charger.
		%Through this virtual photon charging protocol, 
		Owing to the significantly suppressed dissipation,
		%the QB's energy can be charged to a nearly 100$\%$ state and 95$\%$ converted into extractable work, 
		the stored energy of QB can be charged to a nearly full state and the stored energy is almost converted into extractable work, 
		which greatly improves the energy conversion efficiency. 
		%quantum battery can be fully charged and the energy stored in the battery can be fully extracted
		%	Therefore, through our virtual photon charging protocol, the self-discharging effect brought by the charger to the QB can be significantly suppressed, and we predict that the half-life time of the QB can reach $\tau=1.5$ s under current experimental conditions.
	\end{abstract}
	%The  cat state, as an important tool to explore the classical-quantum boundary, not only provides important tests of macroscopic quantum effects, but also has important applications in quantum metrology and quantum computation. The three-magnon process, which originates from the magnetic dipole-dipole interaction,
	%has been less explored in the preparation of non-classical magnon state. 
	%Here, we develop the magnon parametric coupling process that arises from the three-magnon scattering.
	%The magnon parametric coupling strength is analytically derived and can reach a strong coupling regime within a discretized magnon mode system.
	%Based on this, we propose a new scheme to prepare the cat state without involving external nonlinearity in a magnetic system. Through the magnon parametric coupling, the extra magnon dissipation channel provided by the magnon-photon interaction will induce an effective two-magnon loss channel, then steering the magnon into the cat state manifold. 
	%Notably, there is an inverse proportionality between the size of magnon cat states and the parametric coupling strength, which facilitates the preparation of large-size cat states, even under environmental decoherence.
	%Our work opens a new path for the preparation of non-classical states in magnon spintronics systems without extra nonlinear materials, and will also promote the study of fundamental nonlinear physics in the magnetic platform.

	\maketitle
	\textit{Introduction.}--Quantum thermodynamics, as an emerging interdisciplinary field, is receiving significant attention due to its combination of the basic principles of quantum mechanics into the energy and work in thermodynamics \cite{PhysRevLett.115.210403,bera2017generalized,PhysRevLett.112.030602,PhysRevLett.122.110601,micadei2019reversing,mayer2023generalized,guryanova2016thermodynamics}. A fascinating research object in quantum thermodynamics is emerging, quantum battery (QB) \cite{PhysRevE.87.042123,PhysRevLett.111.240401,kzvn-dj7v,doi:10.1126/sciadv.abk3160,PhysRevA.109.012204,PhysRevA.110.062209,PhysRevLett.122.210601,RevModPhys.96.031001,PhysRevLett.134.180401,PhysRevE.108.064106,10.1063/5.0156618,downing2023quantum,PhysRevA.109.052206,Downing_2024,PhysRevLett.132.090401,PhysRevLett.132.210402,67wh-1fxv,hadipour2025amplified,xqtv-qbyk,10.1002/adma.202415073,ferraro2026opportunities,10.1002/qute.202500845,PhysRevLett.124.130601,PhysRevLett.131.030402,PhysRevLett.132.240401,PhysRevA.109.062614,PhysRevResearch.2.023113,PhysRevApplied.14.024092,PhysRevE.101.062114}, an energy storage and transmission device at microscopic scales. Charging power and energy transfer rate of traditional electrochemical batteries are greatly constrained by classical thermodynamic laws, material dependency, and sustainability \cite{p93y-jflt}. On the contrary, 
	%thanks to the support of non-classical quantum resources, such as quantum entanglement, coherence and quantum superposition, 
	QB can achieve superior charging power (collective charging effect) \cite{kzvn-dj7v,PhysRevResearch.6.023136,PhysRevA.110.032211,shastri2025dephasing,PhysRevLett.118.150601,PhysRevLett.127.100601,PhysRevLett.128.140501,PhysRevA.101.032115,PhysRevB.102.245407,PhysRevE.104.024129,PhysRevA.104.032207,PhysRevResearch.6.023136,Binder_2015,qhz8-mvfb}, higher energy transfer rate \cite{PhysRevLett.131.240401}, enhanced extractable work (entropy) \cite{doi:10.1126/sciadv.abk3160,PhysRevA.109.012204,PhysRevA.110.062209,PhysRevE.87.042123,PhysRevResearch.6.023136}, minimized energy loss \cite{PhysRevLett.134.010408} and coherent enhanced energy output by leveraging non-classical quantum resources\cite{RevModPhys.91.025001,PhysRevE.102.052109,PhysRevA.106.062609,PhysRevLett.129.130602} such as quantum superposition, quantum coherence \cite{PhysRevLett.125.040601,RevModPhys.89.041003,PhysRevLett.131.260401,PhysRevA.108.052213,PhysRevLett.125.180603,lostaglio2015description,PhysRevX.5.031044,Korzekwa_2016,PhysRevE.102.042111}, and quantum entanglement \cite{RevModPhys.81.865,PhysRevLett.122.047702,PhysRevB.104.245418,PhysRevA.107.022215,PhysRevA.110.022433,PhysRevX.5.041011,PhysRevLett.89.180402,Allahverdyan_2004}. An important quantum advantage, the collective, the quadratic scaling behavior of charging power has been demonstrated \cite{PhysRevA.110.062209,PhysRevLett.125.236402,PhysRevE.105.064119} in both superconducting quantum circuits \cite{10.1002/qute.202400651} and quantum optical systems \cite{PhysRevLett.120.117702,PhysRevB.105.115405}. So far, a variety of QB models have been explored, ranging from the two-level systems (TLSs) \cite{PhysRevA.106.042601,PhysRevLett.134.130401}, three-level systems \cite{PhysRevE.100.032107,6c73-ll23,dou2022highly,PhysRevB.109.235432}, Dicke model \cite{PhysRevLett.120.117702,PhysRevLett.133.243602,PhysRevB.105.115405,PhysRevA.109.022210}, the Tavis-Cummings model \cite{PhysRevA.109.012204,PhysRevA.109.012204}, Sachdev-Ye-Kitaev model \cite{PhysRevLett.125.236402,PhysRevResearch.4.043150,PhysRevLett.124.244101}, the spin chain model \cite{PhysRevA.97.022106,PhysRevLett.133.197001,PhysRevA.106.032212,PhysRevResearch.4.013172} to the quantum harmonic oscillator \cite{PhysRevA.107.042419,PhysRevB.98.205423,PhysRevB.99.035421,Grebenkov_2020}, the cavity optomechanical model \cite{PhysRevA.110.062204} and magnon-mediated quantum battery \cite{PhysRevA.104.032606}. The rapid development of QB opens new path for future high-density energy storage and power supply for quantum computing \cite{PhysRevX.11.021014,niedenzu2018quantum,bhattacharjee2021quantum,Giorgi_2015}, while also promotes the theoretical framework of quantum thermodynamics and plays important role for future information technology \cite{PhysRevResearch.2.023095}, energy science \cite{PhysRevLett.129.110601,PhysRevLett.131.260401}, and other technologies \cite{parrondo2015thermodynamics,Millen_2016,PRXQuantum.3.020101}.
	
	Although QB show important theoretical advantages over their classical counterparts \cite{horodecki2013fundamental,skrzypczyk2014work}, there are still important problems and challenges. On the one hand, the problem of energy backflow from the QB to the charger is severe, which will %cause the energy of the battery to flow to the charger, 
	result in impaired charging efficiency of the QB and preventing the advantages of quantum batteries. Important schemes to deal with this problem have been achieved, such as the use of optimal pulse control sequences \cite{PhysRevA.107.032218,PhysRevE.106.014138,PhysRevA.110.052601}, non-directional energy transfer \cite{PhysRevLett.132.210402,67wh-1fxv,p93y-jflt}. Another important problem is that inevitable spontaneous loss of QB energy caused by the decoherence of the environment \cite{PhysRevB.99.035421,PhysRevA.110.052404,vqnk-kzqg}, which is also called the self-discharging of the QB \cite{PhysRevE.103.042118,PhysRevE.105.054115,PhysRevE.109.054132}. Two different reasons lead to the self-discharging of the QB, one is the energy loss channel of the QB itself, and the other is the self-discharge channel brought by the charger. Although some remarkable progress has been reported, such as dark state engineering \cite{PhysRevApplied.14.024092}, floquet engineering, quantum reservoir engineering \cite{PhysRevE.104.064143,Kamin_2020,Li:22,PhysRevA.109.012224,PhysRevE.100.032107} and feedback control \cite{PhysRevE.106.014138}. 
		%However, the batteries implemented by these schemes always require additional auxiliary systems, such as special environment libraries. 
		However, these schemes are implemented in ideal cases, such as waveguide libraries without dissipation \cite{PhysRevA.102.060201,PhysRevA.96.043811}. How to simultaneously suppress the self-discharging of the QB and the energy backflow between the QB and the charger is still an important task and  faces challenges.
		
	%Although the quantum battery used in the engineering application of the environment library can store energy significantly, the additional dissipation channel of the environment library exists in practice.
	
\begin{figure}[h]
		\includegraphics[scale=0.8]{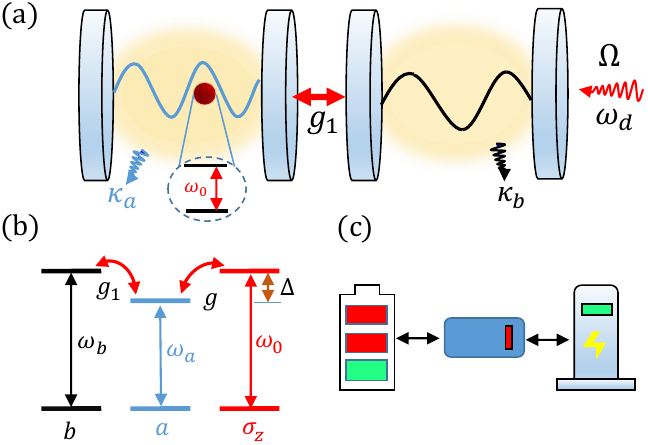}
		\caption{ (a) Schematic diagram of QB consisting of a auxiliary cavity (left) and pump cavity (right). (b) Schematic diagram of energy levels for a near-resonant QB and charger. (c) Schematic diagram of a QB utilizing the virtual photon charging protocol.}\label{fig-01} %The parameters are chosen as (a) $\Delta=10\kappa_a$, $g_1=\kappa_a$, and $g=2\kappa_a$, (b) $\Omega=0.5\kappa_{a}$, and $\kappa_m=0.001\kappa_{a}$.}\label{fig-01}
\end{figure}

	Here, %we propose a virtual photon charging protocol for QB, in which an effective second-order coupling process between the near-resonant QB and the charger is formed through the large detuning auxiliary cavity. 
	we propose a virtual photon charging protocol for QB, in which the auxiliary cavity mediates a virtual-photon coupling between the QB and the charger while strongly suppressing the decay of the QB via Purcell effect. Moreover, Increasing the charger dissipation drives this remaining loss into the quantum Zeno regime, further reducing the effective self-discharging rate of the QB. As a result, the charger-induced decay can be suppressed by four orders of magnitude %yielding a half-life of 1.5s under realistic parameters. 
	%The protected battery can be charged close to unity, with about 95 of the stored energy converted into extractable work.
	%Meanwhile, through the Purcell effect, the additional dissipation rate of the QB caused by the large detuning auxiliary cavity can be significantly suppressed, so the QB 
	%can reach nearly 100$\%$ of the charging degree, and all of it can be converted into ergotropy, 
%	We reveal that the Purcell effect could induce the quantum Zeno phenomenon under the strong coupling regime, in which the greater the dissipation of the charger, the lower the self-discharging rate of the QB. Therefore, our scheme can significantly suppress the self-discharging of the QB caused by the charger four orders of magnitude. 
	Meanwhile, the quantum Zeno effect induced by the Purcell effect can also avoid the energy backflow between the QB and the charger.
	Thanks to the Purcell effect's strong suppression of dissipation,
	the QB can be fully charged, with its stored energy is almost converted into extractable work, 
	%and the 95$\%$ of stored energy in the QB can be converted into extractable work benefiting from the important suppression of dissipation by the Purcell effect,
	which greatly improves the conversion efficiency of the QB.  %by increasing the dissipation rate of the charger and can significantly resist the dissipation caused by the charger. 
	\textit{Virtual photon charging model.}-----The virtual photon charging protocol consists of a two-level atom, an auxiliary cavity and a pump cavity, where the pump cavity and the atom act as the charger and quantum battery (QB) (Fig. \ref{fig-01}). The atom is placed in the auxiliary cavity, and the interaction between the auxiliary cavity and the atom is described by a JC model under the rotating wave approximation. In addition, the auxiliary cavity is coupled to the pump cavity.
	%In our model, a two-level atoms with a transition frequency $\omega_0$ is used as a quantum battery (QB). $\sigma^+=|e\rangle\langle g|$ is the rising operator of the atom, and $\sigma^-=|g\rangle\langle e|$ is the lowering operator of the atom. $|e\rangle$ is the excited state of the atom, and $|g\rangle$ is the ground state of the atom. Two photon cavities acts as a charger. 
	%By placing the atom in the auxiliary cavity, the interaction between the auxiliary cavity and the atoms is described by a JC model under the rotating wave approximation. In addition, the auxiliary cavity is coupled to the pump cavity. 
	We consider the pump cavity is driven by a coherent laser with driving strength $\Omega$, and frequency $\omega_d$, and the total Hamiltonian of the system is written as  
	\begin{equation}
		\begin{aligned}
			H=&\omega_aa^\dagger a+\omega_0\sigma^+\sigma^-+\omega_bb^\dagger b+g(a\sigma^++a^\dagger \sigma^-)\\
			&+g_1(ab^\dagger+a^{\dagger} b)+\Omega(be^{i\omega_dt}+b^\dagger e^{-i\omega_dt}),
		\end{aligned}\label{eq-000}
	\end{equation}
	where $a$ is the annihilation operator of auxiliary cavity with frequency $\omega_a$ and $b$ is the annihilation operator of pump cavity with frequency $\omega_b$. 
	%a two-level atoms with a transition frequency $\omega_0$ is used as a quantum battery (QB). 
	$\sigma^+=|e\rangle\langle g|$ is the rising operator of the atom, and $\sigma^-=|g\rangle\langle e|$ is the lowering operator of the atom. $|e\rangle$$(|g\rangle)$ is the excited (ground) state of the atom. $\omega_0$ is the transition frequency from the excited state $|e\rangle$ to the ground state $|g\rangle$.
	%Two photon cavities acts as a charger.
	$g$ describes the interaction strength between the auxiliary cavity and the atom. $g_1$ is the hopping strength between two cavities. %Here, we define $S_+=\sigma^+_1+\sigma^+_2$ and $S_-=\sigma^-_1+\sigma^-_2$.
	We rewrite the Hamiltonian in the interaction picture by assuming that $\omega_d\approx\omega_0\approx\omega_b$
	\begin{equation}
		\begin{aligned}
			H_1=&g(a\sigma^+e^{i\Delta t}+a^\dagger \sigma^-e^{-i\Delta t})
			+g_1(ab^\dagger e^{i\Delta t}+a^{\dagger} be^{-i\Delta t})\\&+\Omega(b+b^\dagger)\label{eq-01},
		\end{aligned}
	\end{equation}
	where $\Delta=\omega_0-\omega_a$ is the detuning of the atom to the auxiliary cavity $a$. 
	
	%\textit{Virtual photon charging process.}-----
	When the auxiliary cavity is greatly detuned with the atom, the auxiliary cavity will be in a vacuum state and will not be excited. Meanwhile, since the atom is near-resonant with the pump cavity, there is an effective coupling between the atom and the pump cavity through the virtual photon process constructed by auxiliary cavity. Specific, when $\Delta\gg\{\Omega,g,g_1\}$, the cavity $a$ is in the vacuum state and can be adiabaticly eliminated through time averaging \cite{PhysRevA.82.052106,PhysRevA.95.032124}, and the effective second-order Hamiltonian is written as \cite{supp}
	\begin{equation}
		H_{\text{eff}}=J(\sigma^+b+\sigma^-b^\dagger)+\Omega(b+b^\dagger),
	\end{equation}
	where the second-order term $\frac{g^2}{\Delta}\sigma^+\sigma^-$ and $\frac{g_1^2}{\Delta}b^\dagger b$ can be eliminated by shifting the eigenfrequencies of the atom and pump cavity. $J=\frac{gg_1}{\Delta}$ is the effective coupling strength between the atom and the pump cavity. Through the pump cavity, we can realize the charging of the QB through the virtual photon process.
%\begin{equation}
%	H_{\text{eff}}^{(2)}=\frac{g\Omega}{\Delta}(\sigma_++\sigma_-),
%\end{equation}

\textit{Self-discharging of the QB.}-----When the QB is charged, its charger and auxiliary cavity will bring additional dissipation processes to the QB. For these dissipations of cavities, we use the master equation with the lindbland operator to describe them, which is written as
\begin{equation}
	\begin{aligned}
		\dot{\rho}=-i[H,\rho]+(\kappa_{a}/2)\mathcal{L}[a]\rho+(\kappa_{b}/2)\mathcal{L}[b]\rho\label{eq-02},
	\end{aligned}
\end{equation}
where $\kappa_{a}$ is the single-photon loss of auxiliary cavity and $\kappa_{b}$ is the single-photon loss of charger, and $\mathcal{L}[o]\rho=2o\rho o^{\dagger }-o^{\dagger } o\rho-\rho o^{\dagger} o$ is a lindblad dissipative operator. The photon dissipation channels will cause the dissipation of atom. 
%\textcolor{red}{However, in our scheme, the atom is large detuned to the auxiliary cavity, %and the atomic dissipation rate will be greatly reduced. 
	%and the atomic dissipation rate brought by the auxiliary cavity will be greatly suppressed by the Purcell effect.}
%According to the formal theory of dissipation, the dissipation of atom caused by photon dissipation channels can be derived and the master equation of the system can be rewritten as 
%\begin{equation}
%	\begin{aligned}
	%		\dot{\rho}=-i[H_{\text{eff}},\rho]+(\kappa_{e}/2)\mathcal{L}[\sigma_-]\rho+(\kappa_{b}/2)\mathcal{L}[b]\rho\label{eq-005},
	%	\end{aligned}
%\end{equation}
%where $\kappa_{e}=\frac{\kappa_{a}g^2}{\Delta^2+\kappa_{a}^2/4}$ is the effective dissipation rate of atom \cite{supp}. It is noted that the dissipation rate of atom is accompanied by the $\frac{g^2}{\Delta^2+\kappa_{a}^2/4}$ factor, which implies the large detuning can suppress the QB self-discharging.
%The Eq. (\ref{eq-005}) can describe the charging and self-discharge process of the QB. 
The Eq. (\ref{eq-02}) can describe the charging and self-discharging process of the QB. When the driving strength $\Omega\neq0$, the external energy is input into the system, and the QB can be charged through the virtual photon process. When the driving strength is $\Omega=0$, we can use the Eq. (\ref{eq-02}) to study the self-discharging effect of the QB. %phenomenon
%The first term presents the charging protocol of virtual photons, and the second term is the dissipation caused by photon dissipation.
%Through Eq. (\ref{eq-05}), we can investigate the performance of QB.

\textit{QB performance indicators.}-----Then, let's discuss the performance of QB. As an energy storage device, its performance is characterized by the stored energy.
The stored energy of the QB is
\begin{equation}
	\begin{aligned}
		E(t)=\text{Tr}[\rho_B(t)H_B],
	\end{aligned}
\end{equation}
where $H_B=\omega_0\sigma^+\sigma^-$ and $\rho_B(t)=\text{Tr}_{a,b}[\rho(t)]$ is the reduced matrix of atom.
Not all the stored energy in the QB can be extracted, and the maximal amount of work that can be extracted form a state $\rho_B(t)$ is provided by the ergotropy
\begin{equation}
	\begin{aligned}
		\xi(t)=\text{Tr}[\rho_B(t)H_B]-\text{Tr}[\bar{\rho}_B(t)H_B],
	\end{aligned}
\end{equation}
where $\bar{\rho}_B(t)=\sum_nr_n(t)|s_n\rangle\langle s_n|$ is the passive state, $r_n(t)$ are the eigenvalues of $\rho_B(t)$ ordered in a
descending sort, and ${|s_n\rangle}$ are the eigenstates of $H_B$ with the corresponding eigenvalues $s_n$ ordered in an ascending sort.
Through some algebras, we can get a simplified ergotropy expression
$\xi(t)=\text{Tr}[\rho_B(t)H_B]-\sum_nr_n(t)s_n$. In addition, the average charging power of QB is given by 
\begin{equation}
	\begin{aligned}
		P(t)=\frac{\xi(t)}{t},
	\end{aligned}
\end{equation}
where describes the speed of QB charging.

%\textit{Charging of the QB.}-----When the battery is charged, its charger will bring additional dissipation process to the battery. For this dissipative process, we use the master equation with the lindbland operator to describe it, which is written as
%\begin{equation}
%	\begin{aligned}
	%		\dot{\rho}=-i[H_e,\rho]+(\kappa_{a}/2)\mathcal{L}[a]\rho+(\kappa_{b}/2)\mathcal{L}[b]\rho\label{eq-02},
	%	\end{aligned}
%\end{equation}
%where $\kappa_{a}$ is the single-photon loss of signal cavity and $\kappa_{b}$ is the single-photon loss of pump cavity, and $\mathcal{L}[o]\rho=2o\rho o^{\dagger }-o^{\dagger } o\rho-\rho o^{\dagger} o$ is a lindblad dissipative operator.
%In the process of charging, QB will inevitably be brought to the self-discharge channel by the charger. In order to investigate the influence of the dissipative channel caused by the charger on the battery, we use the master equation to study the evolution of the system.
%The master equation of the original system is read as
%\begin{equation}
%	\begin{aligned}
	%		\dot{\rho}=-i[H,\rho]+(\kappa_{a}/2)\mathcal{L}[a_1]\rho\label{eq-02},
	%	\end{aligned}
%\end{equation}
%where $\kappa_{a}$ is the single-photon loss, and $\mathcal{L}[o]\rho=2o\rho o^{\dagger }-o^{\dagger } o\rho-\rho o^{\dagger} o$ is a lindblad dissipative operator.

\begin{figure}[bh]
	\includegraphics[scale=0.585]{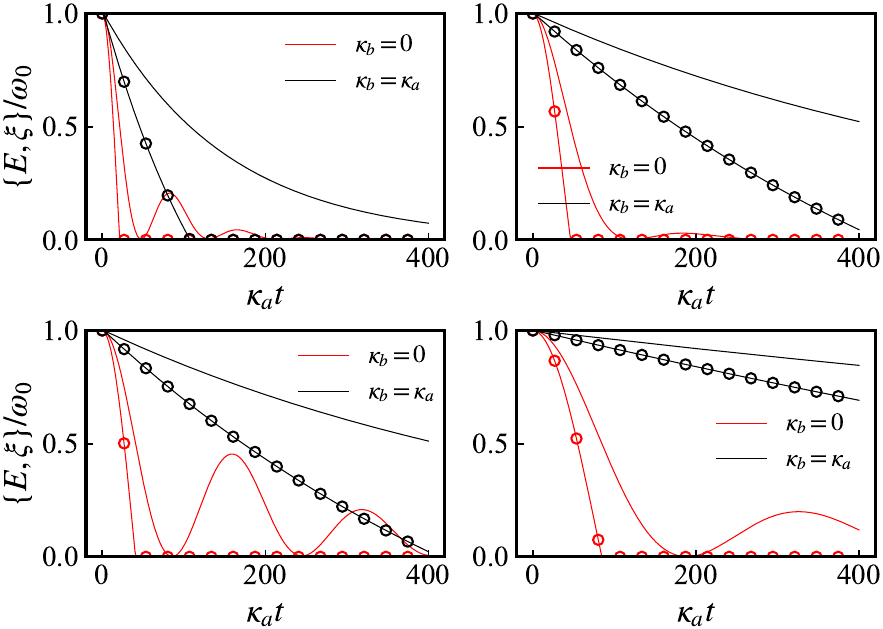}
	\caption{ The stored energy and ergotropy of the QB versus time under different charger dissipation and detuning. The unmarked line is energy, and the marked line is ergotropy. The parameters are chosen as (a) $\Delta=10\kappa_a$, $\Omega=0$, $g=0.2\kappa_a$ and $g_1=2\kappa_a$, (b) $\Delta=10\kappa_a$, $\Omega=0$, $g=0.1\kappa_a$ and $g_1=2\kappa_a$ (c) $\Delta=20\kappa_a$, $\Omega=0$, $g=0.2\kappa_a$ and $g_1=2\kappa_a$and (d) $\Delta=20\kappa_a$, $\Omega=0$, $g=0.1\kappa_a$ and $g_1=2\kappa_a$.}\label{fig-03}
\end{figure}

\textit{Suppressed the self-discharging of the QB.}-----First, we discuss the self-discharging effect of the QB. Here we assume that the QB has been fully charged.
%after being fully charged. When the energy of the QB is charged to the highest, we turn off the driving, which can be approximated as the QB is in a fully charged state. %which is strictly in the excited state. 
Due to the additional dissipation channels brought by the auxiliary cavity and charger, the QB will self-discharging through those channels. 
In Fig. \ref{fig-03}, we plot the stored energy and ergotropy of the QB versus time at different dissipation rates of charger. Here, we set the QB initially in the excited state, $\Omega=0$ and the two cavities in the vacuum state. The unmarked line is the stored energy of the QB, and the marked line is its corresponding ergotropy.
%\textcolor{blue}{An interesting result is that the increase of charger dissipation not only does not accelerate the self-discharging of the QB, on the contrary, the increase of charger dissipation also delays the self-discharging of the QB.} 
%\textcolor{red}{An interesting result is that when the dissipation of the charger increases, it not only does not accelerate the self-discharging of the QB, but on the contrary, an increase in charger dissipation actually slows down the self-discharging behavior of the QB. \textcolor{cyan}{Interestingly, increasing the charger dissipation does not accelerate the QB's self-discharging; rather, it slows it down.}}
Under the parameter conditions of the strong coupling regime we selected, an interesting result is that when the dissipation of the charger increases, the self-discharging effect of the QB not only does not increase correspondingly. On the contrary, when the dissipation of the charger increases, the self-discharging effect of the QB is weakened.
Specifically, for $g=0.2\kappa_a, g_1=2\kappa_a$ and $\Delta=10\kappa_a$, when the dissipation rate of the charger is $\kappa_b=0$, the half-life time of the QB is $\tau=21/\kappa_a$, where the half-life time is defined as the time when the stored energy decays to half of the initial amount of stored energy. Its energy is completely lost after $\kappa_at=200$. However, when the dissipation rate of the charger is $\kappa_b=\kappa_a$, the half-life time of the QB is $\tau=112/\kappa_a$, which is five times as long as that without dissipation. The stored energy will not be completely lost until $\kappa_at=500$. 
The ergotropy exhibits a dissipation behavior similar to that of stored energy.
With the decrease of cavity-atom interaction, this phenomenon becomes more obvious [Figs. \ref{fig-03}(a) and \ref{fig-03}(b)]. 
%This is a counterintuitive %(interesting) 
%physical result. We believe that this is in the case of large detuning, the introduction of pump cavity dissipation will inhibit the self-discharging of the QB. 
%\begin{figure}[th]
%	\includegraphics[scale=0.58]{Fig_5}
%	\caption{ (a) The evolution of the occupancy numbers of different modes controlled by the original Hamiltonian in the absence of dissipation and driving. (b) The time evolution of different modes using both the original Hamiltonian (solid line) and the effective Hamiltonian (dashed line) with driving and dissipation. The parameters are chosen as (a) $\Delta=10\kappa_a$, $\Omega=0$, $g=0.2\kappa_a$ and $g_1=2\kappa_a$, (b) $\Delta=10\kappa_a$, $\Omega=0$, $g=0.1\kappa_a$ and $g_1=2\kappa_a$ (c) $\Delta=20\kappa_a$, $\Omega=0$, $g=0.2\kappa_a$ and $g_1=2\kappa_a$and (d) $\Delta=20\kappa_a$, $\Omega=0$, $g=0.1\kappa_a$ and $g_1=2\kappa_a$.}\label{fig-01}
%\end{figure}

In order to be able to clarify the physical process behind this, we recall the master equation of the whole system
\begin{equation}
	\dot{\rho}=-i[H,\rho]+(\kappa_{a}/2)\mathcal{L}[a]\rho+(\kappa_{b}/2)\mathcal{L}[b]\rho. 
\end{equation}
The above findings suggest that, rather than accelerating self-discharging, an increase in charger dissipation actually decelerates it. This implies that the QB's self-discharging rate would be higher in the absence of charger dissipation. The key problem is how to obtain an actual self-discharging rate of the QB. When the auxiliary cavity is greatly detuned to the atom and the charger, it is approximately in a vacuum state and can be cut off. Correspondingly, the dissipation channel of the auxiliary cavity induces an additional dissipation process attached to the QB and the pump cavity, and its effective master equation can be written as  
\begin{equation}
	\begin{aligned}
		\dot{\rho}=&-i[H_{\text{eff}},\rho]+(\kappa_{e}/2)\mathcal{L}[\sigma_-]\rho+(\kappa_{s}/2)\mathcal{L}[b]\rho\label{eq-09},
	\end{aligned}
\end{equation}
where $\kappa_{e}=\frac{\kappa_{a}g^2}{\Delta^2+\kappa_{a}^2/4}$ is the effective dissipation rate of QB and $\kappa_s=\kappa_b+\kappa_f=\kappa_b+\frac{\kappa_{a}g_1^2}{\Delta^2+\kappa_{a}^2/4}$ is the total dissipation rate of charger \cite{supp}.
$H_{\text{eff}}$ is the effective Hamiltonian between the QB and the charger. 
%The  Eq. 9 shows that the effective dissipation rate $\kappa_e$ of the atom induced by the large detuning auxiliary cavity and the additional dissipation rate $\kappa_f$ of the pump cavity are squared respectively, which is related to the cavity atom coupling rate and the hopping interaction between cavities. Meanwhile, they are inversely proportional to the detuning, suggesting that these rates can be significantly affected by adjusting the detuning and coupling strength. 
Eq. \ref{eq-09} shows that the effective dissipation rate $\kappa_e$ of the QB and the additional dissipation rate $\kappa_f$ of the charger induced by the large-detuning auxiliary cavity are respectively squared related to the cavity-atom coupling strength and the inter-cavity hopping interaction. Meanwhile, they are both inversely proportional to the square of the detuning, suggesting that these rates can be significantly influenced by the detuning and coupling strength.
%We define the total dissipation rate of the pump cavity as $\kappa_s=\kappa_f+\kappa_{b}$. 
When the self-dissipation rate of the charger is $\kappa_b=0$, the total dissipation $\kappa_s$ of the charger is equal to the additional dissipation $\kappa_f$ induced by the auxiliary cavity. For the parameters $g=0.2\kappa_a, g_1=2\kappa_{a}$ and $\Delta=10\kappa_a$ we actually selected, we can calculate the total dissipation rate of charger $\kappa_s\approx4\times10^{-2}\kappa_a$ and the effective dissipation rate of the QB $\kappa_e\approx4\times10^{-4}\kappa_a$, which shows that the dissipation rate of the charger is much larger than that of the QB. In fact, the dissipation is greater than the coupling strength, so this value is only used for analysis. Therefore, the charger will rapidly evolve to a steady state so that it can be adiabatically eliminated. We further obtain the effective master equation of the QB
\begin{equation}
	\begin{aligned}
		\dot{\rho}=&(\kappa_{m}/2)\mathcal{L}[\sigma_-]\rho\label{eq-10},
	\end{aligned}
\end{equation}
where $\kappa_m=\kappa_e+\frac{4g^2g_1^2}{\kappa_s\Delta^2}$ is the actual dissipation rate of the QB. 
%caused by the dissipation of the pump cavity. 
We note that the actual dissipation rate $\kappa_m$ is inversely proportional to the total dissipation rate $\kappa_s$ of the charger, suggesting that the larger the dissipation rate of the charger, the lower the actual dissipation rate of the QB. 
This behavior is similar to the quantum Zeno effect in weakly coupled systems \cite{PhysRevLett.126.190402}. Here, the Purcell effect significantly reduces the dissipation of the QB and charger reduced by auxiliary cavity through the detuning, and the original strong coupling regime is mapped to a weak coupling regime by the detuning. Thus, the stronger the dissipation of the charger, the weaker the self-discharging effect of the QB.
% This behavior is actually the quantum Zeno effect \cite{PhysRevLett.126.190402}, but it occurs in a strongly coupled system induced by the Purcell effect.
%This result shows that a dissipative system with large detuning can present a physical result similar to that induced by the quantum Zeno effect by adjusting the coupling strength. In connection with the quantum Zeno effect, 
This conclusion can be extended to the case where the self-dissipation rate of the charger is not zero, i.e., $\kappa_b\neq0$, because the self-dissipation rate of the charger only increases the total dissipation rate $\kappa_s$ so that the conditions for adiabatic elimination are more satisfied. 
Therefore, our solution can significantly suppress the self-discharging rate of QB brought by the charger, thereby greatly inhibiting QB aging.
For $\Delta=10\kappa_a, g=0.2\kappa_a, g_1=2\kappa_a$, the actual self-discharging rate of QB brought by the charger is $\kappa_m\approx\frac{g^2}{\Delta^2}\kappa_a\sim10^{-4}\kappa_a$, implying the self-discharging rate can be suppressed by four orders of magnitude.
According to the actual dissipation rate $\kappa_m=\kappa_e+\frac{4g^2g_1^2}{\kappa_s\Delta^2}$ given by Eq. 27, we could increase the detuning to greater suppress the self-discharging of the QB. %, which will significantly delay the self-discharging of the QB. 
%Due to the Purcrll effect, the self-discharge rate of the battery will be greatly reduced. 
%Therefore, by increasing the detuning, the self-discharging of the QB will be greater suppressed.
In Figs. \ref{fig-03}(c) and \ref{fig-03}(d), we plot the stored energy and ergotropy of QB versus time under detuning $\Delta=20\kappa_a$. 
%On the one hand, 
With the increase of the detuning, the increase of the dissipation of the charger has a more obvious effect on the self-discharging suppression of the QB. %On the other hand, 
%Due to the increase of detuning, 
For $\Delta=20\kappa_a, g=0.1\kappa_a, g_1=2\kappa_a$, 
the self-discharging half-life time of QB is greatly extended from $\tau=82/\kappa_a$ to $\tau=1500/\kappa_a$ (Fig. \ref{fig-03}(d)). Correspondingly, the decay of ergotropy is also significantly suppressed (marked lines).
%We can see that with the increase of the detuning, the half-life time of the QB is prolonged. In particular, for the detuning $\Delta=20\kappa_a$, the half-life time of the QB is $\tau=100$. 
%As the detuning increases to $\Delta=20\kappa_a$, the half-life time of the battery further increases to $\tau=200$.
Considering the realistic cavity with high quality factor, the single-photon dissipation rate can reach to $\kappa_a=1$ kHZ, which corresponds to the half-life time of QB $\tau=1.5$ s, indicating that the half-life time of the QB can be increased by four orders of magnitude. At the same time, we also note that our scheme can also avoid energy reflux between the battery and the charger.

\begin{figure}[h]
\includegraphics[scale=0.635]{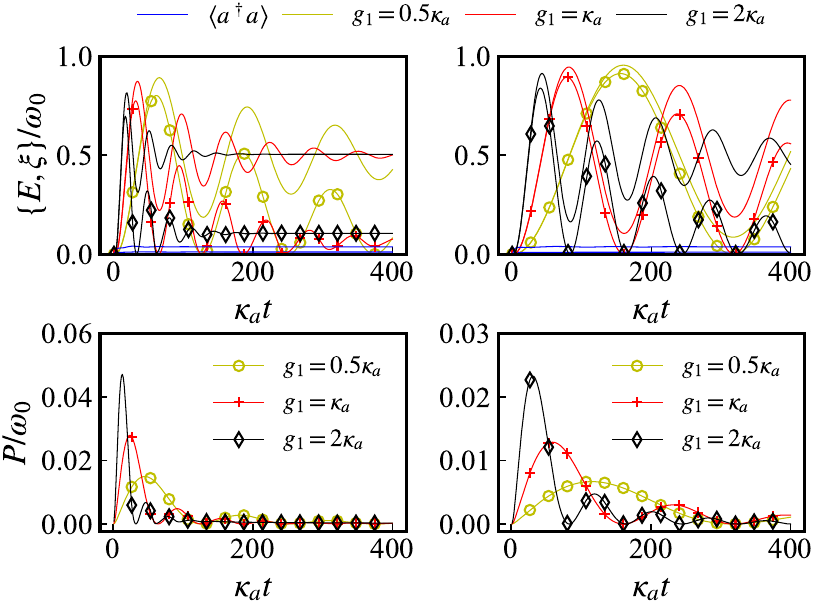}
\caption{ (a) and (b) The stored energy and ergotropy of the QB versus time under different hopping strengths $g_1$ and cavity-atom interaction strengths $g$. The unmarked line is stored energy, and the marked line is ergotropy. (c) and (d) The average charging power of the QB versus time under different hopping strengths and cavity-atom interaction strengths. The parameters are chosen as $\Delta=10\kappa_a$, $\Omega=0.5\kappa_{a}$ and (a) $g=0.5\kappa_a$, (b) $g=0.2\kappa_a$, (c) $g=0.5\kappa_a$, (d) $g=0.2\kappa_a$.}\label{fig-02}
\end{figure}

\textit{Charging of the QB.}-----Then we discuss the charging of the QB. When we charge the QB, the external driving on the pump cavity is turned on, and the energy is transmitted from the charger to the QB to complete the charging of the QB. 
%As a large detuning charging protocol, whether the QB can be charged is a very important indicator. 
In Fig. \ref{fig-02}(a), we plot the stored energy $E(t)$ of the QB and the ergotropy $\xi(t)$ versus time. Here, we set the QB to be initially in the ground state, $\Omega\neq0$ and the two cavities are in a vacuum state. From Fig. \ref{fig-02}(a), we can see the QB is charged, i.e., $E_m=0.89$ at $\kappa_at=65$. %and then gradually decreases due to the existence of dissipation.
Therefore, the QB can achieve charging through our charging protocol.
For the ergotropy $\xi(t)$ (mark in Fig. \ref{fig-02}(a)), 
%its evolution is similar to the stored energy $E(t)$. Through the charging protocol of virtual photons, the ergotropy gradually rises to the maximum. 
we note that the ergotropy is lower than the stored energy of the QB, i.e., $\xi_m=0.79$ at $\kappa_at=61$.
Here, we define the charging peak stored energy $E_m$ and peak ergotropy $\xi_m$ as the maximum value of the QB energy during the charging process.
%Let us discuss the effect of interaction $g$ between QB and auxiliary cavity on the performance of QB charging. 
%From Figs. \ref{fig-02}(a) to \ref{fig-02}(b), 
%With the decrease of $g$ (Figs. \ref{fig-02}(a) to \ref{fig-02}(b)), the peak stored energy and ergotropy of the QB increase to $E_m=0.96$ and $\xi_m=0.91$ from $E_m=0.89$ and $\xi_m=0.79$ for $g_1=0.5\kappa_a$, suggesting that the QB has been fully charged.
With the decrease of $g$ (Figs. \ref{fig-02}(a) to \ref{fig-02}(b)), the peak stored energy of the QB increase to $E_m=0.96$ from $E_m=0.89$ for $g_1=0.5\kappa_a$, suggesting that the QB has been fully charged.  
Meanwhile, the peak ergotropy of the QB also increase to $\xi_m=0.91$ from $\xi_m=0.79$. 
%It is not only an increase in quantity, but also an increase in efficiency.
%This not only indicates that the QB has been fully charged, but also significantly increases the ergotropy. Compared with before, 
By defining the extractable efficiency $\eta$ as the ratio of the peak ergotropy to the total stored energy, the extractable efficiency of the QB is also improved, i.e., from $\eta=0.89$ to $\eta=0.95$, suggesting that the QB is almost all extractable work.
%By calculation, the ratio of extractable work to total stored energy ${\xi_m}/{E_t}$ increases from $0.88$ to 0.94, suggesting that the extraction efficiency of the QB is improved. %suggesting that the QB is almost all extractable work. 
	%The physical mechanism is that the dissipation rate brought by the auxiliary cavity to the QB is significantly suppressed by the detuning, i.e.,
	The increase in ergotropy is attributed to the significant suppression of the QB's dissipation rate by the detuning $\kappa_e=\frac{\kappa_{a}g^2}{\Delta^2+\kappa_{a}^2/4}$. % and the Zeno-like effect, 
%so we can approximately assume that the QB has been in a pure state during the evolution process, resulting in extractable work is equal to the total energy \cite{supp}.}
This shows the important advantages of our QB.
Therefore, in order to achieve fully charging and more extractable work of the QB, we can reduce the interaction between the auxiliary cavity and the QB. This inverse relationship can greatly enhance the application of the QB. Of course, the improvement of peak stored energy and ergotropy will also bring some disadvantages, such as the increase of charging time of QB. 
The rate of QB charging is characterized by charging power $P(t)$. In Figs. \ref{fig-02}(c) and \ref{fig-02}(d), we see the charging power $P$ is reduced by half, resulting in a longer charging time of the QB. Therefore, in practical applications, we need to combine these two aspects to select the most suitable coupling strength $g$.
%which imply the stored energy of QB is fully converted into extractable work. 
%\textcolor{red}{
%%The physical mechanism is that the dissipation rate brought by the auxiliary cavity to the QB is significantly suppressed by the detuning, i.e.,
%The physical mechanism is that the dissipation rate of the QB is significantly suppressed by the detuning $\kappa_e=\frac{\kappa_{a}g^2}{\Delta^2+\kappa_{a}^2/4}$, % and the Zeno-like effect, 
%so we can approximately assume that the QB has been in a pure state during the evolution process, resulting in extractable work is equal to the total energy \cite{supp}.} %is a pure state evolution
%resulting in the QB is approximately in a pure state.}
%resulting in a decrease in QB energy at the same rate as the disappearance of coherence.}
%so that the rate of reduction of the QB energy is consistent with the rate of coherence reduction, so that the ergotropy is always equal to the stored energy.
%which means that the energy of the QB is a 100$\%$ utilization efficiency. 
At the same time, in order to show the effectiveness of our second-order Hamiltonian, we use the full Hamiltonian $H$ to plot the evolution of the $\langle a^\dagger a\rangle$ versus time, as shown in the blue lines in the figure. It has been in a vacuum state so it can confirm the validity of our second-order process.
%It is consistent with the time evolution of effective Hamiltonian $H_e$ control, which confirms the effectiveness of our second-order process.

%\begin{figure}[h]
%	\includegraphics[scale=0.635]{reFigure_3}
%	\caption{ (a) The evolution of the occupancy numbers of different modes controlled by the original Hamiltonian in the absence of dissipation and driving. (b) The time evolution of different modes using both the original Hamiltonian (solid line) and the effective Hamiltonian (dashed line) with driving and dissipation. The parameters are chosen as (a) $\Delta=10\kappa_a$, $\Omega=0.5\kappa_{a}$ and $g=0.5\kappa_a$, (b) $g=0.2\kappa_a$, (c) $g=0.5\kappa_a$, (d) $g=0.2\kappa_a$.}\label{fig-02}
%\end{figure}
%The parameter coupling strength will significantly affect the performance of quantum batteries. 

The hopping strength $g_1$ between the cavities will significantly affect the charging performance of the QB.
In Fig. \ref{fig-02}(a), we investigate the stored energy $E(t)$ and the ergotropy $\xi(t)$ of the QB versus time under different hopping strengths. For a fixed cavity-atom interaction $g=0.5\kappa_a$,
%When the hopping strength is $g_1=\kappa_a$, %the maximum energy of the QB can reach 0.8, 
the charging peak stored energy of the QB is $E_m=0.89$ and the peak ergotropy is $\xi_m=0.79$ when the hopping strength is small, i.e., $g_1=0.5\kappa_a$. When the hopping strength is increased to $g_1=\kappa_a$, the peak stored energy and ergotropy of the QB is reduced to $E_m=0.87$ and $\xi_m=0.77$. By further increasing the hopping strength to $g_1=2\kappa_a$, the peak stored energy and ergotropy of the QB reach $E_m=0.81$ and $\xi_m=0.69$, which implies that the QB is not fully charged. The inverse relationship between the hopping strength and the peak stored energy and ergotropy implies that our QB charging scheme has a loose requirement for the inter-cavity hopping strength, which can greatly improve the feasibility of QB applications.
Although the peak stored energy and ergotropy decrease with the increase of the hopping strength, we noticed that the time to reach the peak of the QB was also significantly advanced, for example from $\kappa_at=65$ to $\kappa_at=18$ (from yellow curve to black curve). In Fig. \ref{fig-02}(c), we plot the charging power $P(t)$ of the QB versus time under the corresponding hopping strength. When the hopping strength increases, the charging power of the QB is higher. %which directly reflects the faster charging speed of the QB. 
We can understand that when the hopping strength increases, the effective coupling between the QB and the charger will also increase, resulting in faster energy flowing into the QB.  %to complete charging.
The trade-off between the stored energy and the charging rate of the QB with the hopping strength needs to be better balanced in order to obtain a suitable QB.

\textit{Conclusion}-----Here, we propose a virtual photon QB charging protocol. Through the large detuning auxiliary cavity, an effective second-order coupling between the near-resonant QB and the charger is formed by virtual photon process. 
%At the same time, the self-discharging rate of the QB caused by the large detuning auxiliary cavity will be significantly suppressed by the detuning amount. Therefore, our virtual photon charging protocol can significantly resist the dissipation effect caused by the charger. 
We show that this large detuning charging protocol could exhibit a quantum Zeno phenomenon induced by the Purcell effect, in which the greater the dissipation of the charger, the lower the self-discharging rate of the QB. Therefore, in our scheme, the dissipation caused by the charger to the QB can be suppressed four orders of magnitude. Meanwhile, the quantum Zeno effect induced by the Purcell effect can also avoid the energy reflux between the QB and the charger.
Due to the Purcell effect on the significant suppression of dissipation, the QB can be fully charged and stored energy is almost converted into extractable work, 
which significantly improves the QB's efficiency.

\textit{Acknowledgments.}---We thank Wen Huang for technical support and helpful discussions.
This work was supported by the National Key Research and Development Program of China (2021YFA1400702);
National Natural Science Foundation of China (11975103). The computation is completed in the HPC Platform of Huazhong University of Science and Technology.

%\bibliography{sample}

\begin{thebibliography}{126}%
	\makeatletter
	\providecommand \@ifxundefined [1]{%
		\@ifx{#1\undefined}
	}%
	\providecommand \@ifnum [1]{%
		\ifnum #1\expandafter \@firstoftwo
		\else \expandafter \@secondoftwo
		\fi
	}%
	\providecommand \@ifx [1]{%
		\ifx #1\expandafter \@firstoftwo
		\else \expandafter \@secondoftwo
		\fi
	}%
	\providecommand \natexlab [1]{#1}%
	\providecommand \enquote  [1]{``#1''}%
	\providecommand \bibnamefont  [1]{#1}%
	\providecommand \bibfnamefont [1]{#1}%
	\providecommand \citenamefont [1]{#1}%
	\providecommand \href@noop [0]{\@secondoftwo}%
	\providecommand \href [0]{\begingroup \@sanitize@url \@href}%
	\providecommand \@href[1]{\@@startlink{#1}\@@href}%
	\providecommand \@@href[1]{\endgroup#1\@@endlink}%
	\providecommand \@sanitize@url [0]{\catcode `\\12\catcode `\$12\catcode
		`\&12\catcode `\#12\catcode `\^12\catcode `\_12\catcode `\%12\relax}%
	\providecommand \@@startlink[1]{}%
	\providecommand \@@endlink[0]{}%
	\providecommand \url  [0]{\begingroup\@sanitize@url \@url }%
	\providecommand \@url [1]{\endgroup\@href {#1}{\urlprefix }}%
	\providecommand \urlprefix  [0]{URL }%
	\providecommand \Eprint [0]{\href }%
	\providecommand \doibase [0]{http://dx.doi.org/}%
	\providecommand \selectlanguage [0]{\@gobble}%
	\providecommand \bibinfo  [0]{\@secondoftwo}%
	\providecommand \bibfield  [0]{\@secondoftwo}%
	\providecommand \translation [1]{[#1]}%
	\providecommand \BibitemOpen [0]{}%
	\providecommand \bibitemStop [0]{}%
	\providecommand \bibitemNoStop [0]{.\EOS\space}%
	\providecommand \EOS [0]{\spacefactor3000\relax}%
	\providecommand \BibitemShut  [1]{\csname bibitem#1\endcsname}%
	\let\auto@bib@innerbib\@empty
	%</preamble>
	\bibitem [{\citenamefont {\ifmmode \acute{C}\else
			\'{C}\fi{}wikli\ifmmode~\acute{n}\else \'{n}\fi{}ski}\ \emph
		{et~al.}(2015)\citenamefont {\ifmmode \acute{C}\else
			\'{C}\fi{}wikli\ifmmode~\acute{n}\else \'{n}\fi{}ski}, \citenamefont
		{Studzi\ifmmode~\acute{n}\else \'{n}\fi{}ski}, \citenamefont {Horodecki},\
		and\ \citenamefont {Oppenheim}}]{PhysRevLett.115.210403}%
	\BibitemOpen
	\bibfield  {author} {\bibinfo {author} {\bibfnamefont {P.}~\bibnamefont
			{\ifmmode \acute{C}\else \'{C}\fi{}wikli\ifmmode~\acute{n}\else
				\'{n}\fi{}ski}}, \bibinfo {author} {\bibfnamefont {M.}~\bibnamefont
			{Studzi\ifmmode~\acute{n}\else \'{n}\fi{}ski}}, \bibinfo {author}
		{\bibfnamefont {M.}~\bibnamefont {Horodecki}}, \ and\ \bibinfo {author}
		{\bibfnamefont {J.}~\bibnamefont {Oppenheim}},\ }\bibinfo {title}
	{Limitations on the Evolution of Quantum Coherences: Towards Fully Quantum
		Second Laws of Thermodynamics},\ \href {\doibase
		10.1103/PhysRevLett.115.210403} {\bibfield  {journal} {\bibinfo  {journal}
			{Phys. Rev. Lett.}\ }\textbf {\bibinfo {volume} {115}},\ \bibinfo {pages}
		{210403} (\bibinfo {year} {2015})}\BibitemShut {NoStop}%
	\bibitem [{\citenamefont {Bera}\ \emph {et~al.}(2017)\citenamefont {Bera},
		\citenamefont {Riera}, \citenamefont {Lewenstein},\ and\ \citenamefont
		{Winter}}]{bera2017generalized}%
	\BibitemOpen
	\bibfield  {author} {\bibinfo {author} {\bibfnamefont {M.~N.}\ \bibnamefont
			{Bera}}, \bibinfo {author} {\bibfnamefont {A.}~\bibnamefont {Riera}},
		\bibinfo {author} {\bibfnamefont {M.}~\bibnamefont {Lewenstein}}, \ and\
		\bibinfo {author} {\bibfnamefont {A.}~\bibnamefont {Winter}},\ }\bibinfo
	{title} {Generalized laws of thermodynamics in the presence of
		correlations},\ \href@noop {} {\bibfield  {journal} {\bibinfo  {journal}
			{Nature communications}\ }\textbf {\bibinfo {volume} {8}},\ \bibinfo {pages}
		{2180} (\bibinfo {year} {2017})}\BibitemShut {NoStop}%
	\bibitem [{\citenamefont {Ro\ss{}nagel}\ \emph {et~al.}(2014)\citenamefont
		{Ro\ss{}nagel}, \citenamefont {Abah}, \citenamefont {Schmidt-Kaler},
		\citenamefont {Singer},\ and\ \citenamefont {Lutz}}]{PhysRevLett.112.030602}%
	\BibitemOpen
	\bibfield  {author} {\bibinfo {author} {\bibfnamefont {J.}~\bibnamefont
			{Ro\ss{}nagel}}, \bibinfo {author} {\bibfnamefont {O.}~\bibnamefont {Abah}},
		\bibinfo {author} {\bibfnamefont {F.}~\bibnamefont {Schmidt-Kaler}}, \bibinfo
		{author} {\bibfnamefont {K.}~\bibnamefont {Singer}}, \ and\ \bibinfo {author}
		{\bibfnamefont {E.}~\bibnamefont {Lutz}},\ }\bibinfo {title} {Nanoscale Heat
		Engine Beyond the Carnot Limit},\ \href {\doibase
		10.1103/PhysRevLett.112.030602} {\bibfield  {journal} {\bibinfo  {journal}
			{Phys. Rev. Lett.}\ }\textbf {\bibinfo {volume} {112}},\ \bibinfo {pages}
		{030602} (\bibinfo {year} {2014})}\BibitemShut {NoStop}%
	\bibitem [{\citenamefont {Klatzow}\ \emph {et~al.}(2019)\citenamefont
		{Klatzow}, \citenamefont {Becker}, \citenamefont {Ledingham}, \citenamefont
		{Weinzetl}, \citenamefont {Kaczmarek}, \citenamefont {Saunders},
		\citenamefont {Nunn}, \citenamefont {Walmsley}, \citenamefont {Uzdin},\ and\
		\citenamefont {Poem}}]{PhysRevLett.122.110601}%
	\BibitemOpen
	\bibfield  {author} {\bibinfo {author} {\bibfnamefont {J.}~\bibnamefont
			{Klatzow}}, \bibinfo {author} {\bibfnamefont {J.~N.}\ \bibnamefont {Becker}},
		\bibinfo {author} {\bibfnamefont {P.~M.}\ \bibnamefont {Ledingham}}, \bibinfo
		{author} {\bibfnamefont {C.}~\bibnamefont {Weinzetl}}, \bibinfo {author}
		{\bibfnamefont {K.~T.}\ \bibnamefont {Kaczmarek}}, \bibinfo {author}
		{\bibfnamefont {D.~J.}\ \bibnamefont {Saunders}}, \bibinfo {author}
		{\bibfnamefont {J.}~\bibnamefont {Nunn}}, \bibinfo {author} {\bibfnamefont
			{I.~A.}\ \bibnamefont {Walmsley}}, \bibinfo {author} {\bibfnamefont
			{R.}~\bibnamefont {Uzdin}}, \ and\ \bibinfo {author} {\bibfnamefont
			{E.}~\bibnamefont {Poem}},\ }\bibinfo {title} {Experimental Demonstration of
		Quantum Effects in the Operation of Microscopic Heat Engines},\ \href
	{\doibase 10.1103/PhysRevLett.122.110601} {\bibfield  {journal} {\bibinfo
			{journal} {Phys. Rev. Lett.}\ }\textbf {\bibinfo {volume} {122}},\ \bibinfo
		{pages} {110601} (\bibinfo {year} {2019})}\BibitemShut {NoStop}%
	\bibitem [{\citenamefont {Micadei}\ \emph {et~al.}(2019)\citenamefont
		{Micadei}, \citenamefont {Peterson}, \citenamefont {Souza}, \citenamefont
		{Sarthour}, \citenamefont {Oliveira}, \citenamefont {Landi}, \citenamefont
		{Batalh{\~a}o}, \citenamefont {Serra},\ and\ \citenamefont
		{Lutz}}]{micadei2019reversing}%
	\BibitemOpen
	\bibfield  {author} {\bibinfo {author} {\bibfnamefont {K.}~\bibnamefont
			{Micadei}}, \bibinfo {author} {\bibfnamefont {J.~P.}\ \bibnamefont
			{Peterson}}, \bibinfo {author} {\bibfnamefont {A.~M.}\ \bibnamefont {Souza}},
		\bibinfo {author} {\bibfnamefont {R.~S.}\ \bibnamefont {Sarthour}}, \bibinfo
		{author} {\bibfnamefont {I.~S.}\ \bibnamefont {Oliveira}}, \bibinfo {author}
		{\bibfnamefont {G.~T.}\ \bibnamefont {Landi}}, \bibinfo {author}
		{\bibfnamefont {T.~B.}\ \bibnamefont {Batalh{\~a}o}}, \bibinfo {author}
		{\bibfnamefont {R.~M.}\ \bibnamefont {Serra}}, \ and\ \bibinfo {author}
		{\bibfnamefont {E.}~\bibnamefont {Lutz}},\ }\bibinfo {title} {Reversing the
		direction of heat flow using quantum correlations},\ \href@noop {} {\bibfield
		{journal} {\bibinfo  {journal} {Nature communications}\ }\textbf {\bibinfo
			{volume} {10}},\ \bibinfo {pages} {2456} (\bibinfo {year}
		{2019})}\BibitemShut {NoStop}%
	\bibitem [{\citenamefont {Mayer}\ \emph {et~al.}(2023)\citenamefont {Mayer},
		\citenamefont {Lutz},\ and\ \citenamefont {Widera}}]{mayer2023generalized}%
	\BibitemOpen
	\bibfield  {author} {\bibinfo {author} {\bibfnamefont {D.}~\bibnamefont
			{Mayer}}, \bibinfo {author} {\bibfnamefont {E.}~\bibnamefont {Lutz}}, \ and\
		\bibinfo {author} {\bibfnamefont {A.}~\bibnamefont {Widera}},\ }\bibinfo
	{title} {Generalized Clausius inequalities in a nonequilibrium cold-atom
		system},\ \href@noop {} {\bibfield  {journal} {\bibinfo  {journal}
			{Communications Physics}\ }\textbf {\bibinfo {volume} {6}},\ \bibinfo {pages}
		{61} (\bibinfo {year} {2023})}\BibitemShut {NoStop}%
	\bibitem [{\citenamefont {Guryanova}\ \emph {et~al.}(2016)\citenamefont
		{Guryanova}, \citenamefont {Popescu}, \citenamefont {Short}, \citenamefont
		{Silva},\ and\ \citenamefont {Skrzypczyk}}]{guryanova2016thermodynamics}%
	\BibitemOpen
	\bibfield  {author} {\bibinfo {author} {\bibfnamefont {Y.}~\bibnamefont
			{Guryanova}}, \bibinfo {author} {\bibfnamefont {S.}~\bibnamefont {Popescu}},
		\bibinfo {author} {\bibfnamefont {A.~J.}\ \bibnamefont {Short}}, \bibinfo
		{author} {\bibfnamefont {R.}~\bibnamefont {Silva}}, \ and\ \bibinfo {author}
		{\bibfnamefont {P.}~\bibnamefont {Skrzypczyk}},\ }\bibinfo {title}
	{Thermodynamics of quantum systems with multiple conserved quantities},\
	\href@noop {} {\bibfield  {journal} {\bibinfo  {journal} {Nature
				communications}\ }\textbf {\bibinfo {volume} {7}},\ \bibinfo {pages} {12049}
		(\bibinfo {year} {2016})}\BibitemShut {NoStop}%
	\bibitem [{\citenamefont {Alicki}\ and\ \citenamefont
		{Fannes}(2013)}]{PhysRevE.87.042123}%
	\BibitemOpen
	\bibfield  {author} {\bibinfo {author} {\bibfnamefont {R.}~\bibnamefont
			{Alicki}}\ and\ \bibinfo {author} {\bibfnamefont {M.}~\bibnamefont
			{Fannes}},\ }\bibinfo {title} {Entanglement boost for extractable work from
		ensembles of quantum batteries},\ \href {\doibase 10.1103/PhysRevE.87.042123}
	{\bibfield  {journal} {\bibinfo  {journal} {Phys. Rev. E}\ }\textbf {\bibinfo
			{volume} {87}},\ \bibinfo {pages} {042123} (\bibinfo {year}
		{2013})}\BibitemShut {NoStop}%
	\bibitem [{\citenamefont {Hovhannisyan}\ \emph {et~al.}(2013)\citenamefont
		{Hovhannisyan}, \citenamefont {Perarnau-Llobet}, \citenamefont {Huber},\ and\
		\citenamefont {Ac\'{\i}n}}]{PhysRevLett.111.240401}%
	\BibitemOpen
	\bibfield  {author} {\bibinfo {author} {\bibfnamefont {K.~V.}\ \bibnamefont
			{Hovhannisyan}}, \bibinfo {author} {\bibfnamefont {M.}~\bibnamefont
			{Perarnau-Llobet}}, \bibinfo {author} {\bibfnamefont {M.}~\bibnamefont
			{Huber}}, \ and\ \bibinfo {author} {\bibfnamefont {A.}~\bibnamefont
			{Ac\'{\i}n}},\ }\bibinfo {title} {Entanglement Generation is Not Necessary
		for Optimal Work Extraction},\ \href {\doibase
		10.1103/PhysRevLett.111.240401} {\bibfield  {journal} {\bibinfo  {journal}
			{Phys. Rev. Lett.}\ }\textbf {\bibinfo {volume} {111}},\ \bibinfo {pages}
		{240401} (\bibinfo {year} {2013})}\BibitemShut {NoStop}%
	\bibitem [{\citenamefont {Andolina}\ \emph {et~al.}(2025)\citenamefont
		{Andolina}, \citenamefont {Stanzione}, \citenamefont {Giovannetti},\ and\
		\citenamefont {Polini}}]{kzvn-dj7v}%
	\BibitemOpen
	\bibfield  {author} {\bibinfo {author} {\bibfnamefont {G.~M.}\ \bibnamefont
			{Andolina}}, \bibinfo {author} {\bibfnamefont {V.}~\bibnamefont {Stanzione}},
		\bibinfo {author} {\bibfnamefont {V.}~\bibnamefont {Giovannetti}}, \ and\
		\bibinfo {author} {\bibfnamefont {M.}~\bibnamefont {Polini}},\ }\bibinfo
	{title} {Genuine Quantum Advantage in Anharmonic Bosonic Quantum Batteries},\
	\href {\doibase 10.1103/kzvn-dj7v} {\bibfield  {journal} {\bibinfo  {journal}
			{Phys. Rev. Lett.}\ }\textbf {\bibinfo {volume} {134}},\ \bibinfo {pages}
		{240403} (\bibinfo {year} {2025})}\BibitemShut {NoStop}%
	\bibitem [{\citenamefont {Quach}\ \emph {et~al.}(2022)\citenamefont {Quach},
		\citenamefont {McGhee}, \citenamefont {Ganzer}, \citenamefont {Rouse},
		\citenamefont {Lovett}, \citenamefont {Gauger}, \citenamefont {Keeling},
		\citenamefont {Cerullo}, \citenamefont {Lidzey},\ and\ \citenamefont
		{Virgili}}]{doi:10.1126/sciadv.abk3160}%
	\BibitemOpen
	\bibfield  {author} {\bibinfo {author} {\bibfnamefont {J.~Q.}\ \bibnamefont
			{Quach}}, \bibinfo {author} {\bibfnamefont {K.~E.}\ \bibnamefont {McGhee}},
		\bibinfo {author} {\bibfnamefont {L.}~\bibnamefont {Ganzer}}, \bibinfo
		{author} {\bibfnamefont {D.~M.}\ \bibnamefont {Rouse}}, \bibinfo {author}
		{\bibfnamefont {B.~W.}\ \bibnamefont {Lovett}}, \bibinfo {author}
		{\bibfnamefont {E.~M.}\ \bibnamefont {Gauger}}, \bibinfo {author}
		{\bibfnamefont {J.}~\bibnamefont {Keeling}}, \bibinfo {author} {\bibfnamefont
			{G.}~\bibnamefont {Cerullo}}, \bibinfo {author} {\bibfnamefont {D.~G.}\
			\bibnamefont {Lidzey}}, \ and\ \bibinfo {author} {\bibfnamefont
			{T.}~\bibnamefont {Virgili}},\ }\bibinfo {title} {Superabsorption in an
		organic microcavity: Toward a quantum battery},\ \href {\doibase
		10.1126/sciadv.abk3160} {\bibfield  {journal} {\bibinfo  {journal} {Science
				Advances}\ }\textbf {\bibinfo {volume} {8}},\ \bibinfo {pages} {eabk3160}
		(\bibinfo {year} {2022})}\BibitemShut {NoStop}%
	\bibitem [{\citenamefont {Yang}\ \emph
		{et~al.}(2024{\natexlab{a}})\citenamefont {Yang}, \citenamefont {Shi},
		\citenamefont {Wan}, \citenamefont {Zhang}, \citenamefont {Wang},\ and\
		\citenamefont {Yang}}]{PhysRevA.109.012204}%
	\BibitemOpen
	\bibfield  {author} {\bibinfo {author} {\bibfnamefont {H.-Y.}\ \bibnamefont
			{Yang}}, \bibinfo {author} {\bibfnamefont {H.-L.}\ \bibnamefont {Shi}},
		\bibinfo {author} {\bibfnamefont {Q.-K.}\ \bibnamefont {Wan}}, \bibinfo
		{author} {\bibfnamefont {K.}~\bibnamefont {Zhang}}, \bibinfo {author}
		{\bibfnamefont {X.-H.}\ \bibnamefont {Wang}}, \ and\ \bibinfo {author}
		{\bibfnamefont {W.-L.}\ \bibnamefont {Yang}},\ }\bibinfo {title} {Optimal
		energy storage in the Tavis-Cummings quantum battery},\ \href {\doibase
		10.1103/PhysRevA.109.012204} {\bibfield  {journal} {\bibinfo  {journal}
			{Phys. Rev. A}\ }\textbf {\bibinfo {volume} {109}},\ \bibinfo {pages}
		{012204} (\bibinfo {year} {2024}{\natexlab{a}})}\BibitemShut {NoStop}%
	\bibitem [{\citenamefont {Francica}(2024)}]{PhysRevA.110.062209}%
	\BibitemOpen
	\bibfield  {author} {\bibinfo {author} {\bibfnamefont {G.}~\bibnamefont
			{Francica}},\ }\bibinfo {title} {Quantum advantage in batteries for
		Sachdev-Ye-Kitaev interactions},\ \href {\doibase
		10.1103/PhysRevA.110.062209} {\bibfield  {journal} {\bibinfo  {journal}
			{Phys. Rev. A}\ }\textbf {\bibinfo {volume} {110}},\ \bibinfo {pages}
		{062209} (\bibinfo {year} {2024})}\BibitemShut {NoStop}%
	\bibitem [{\citenamefont {Barra}(2019)}]{PhysRevLett.122.210601}%
	\BibitemOpen
	\bibfield  {author} {\bibinfo {author} {\bibfnamefont {F.}~\bibnamefont
			{Barra}},\ }\bibinfo {title} {Dissipative Charging of a Quantum Battery},\
	\href {\doibase 10.1103/PhysRevLett.122.210601} {\bibfield  {journal}
		{\bibinfo  {journal} {Phys. Rev. Lett.}\ }\textbf {\bibinfo {volume} {122}},\
		\bibinfo {pages} {210601} (\bibinfo {year} {2019})}\BibitemShut {NoStop}%
	\bibitem [{\citenamefont {Campaioli}\ \emph {et~al.}(2024)\citenamefont
		{Campaioli}, \citenamefont {Gherardini}, \citenamefont {Quach}, \citenamefont
		{Polini},\ and\ \citenamefont {Andolina}}]{RevModPhys.96.031001}%
	\BibitemOpen
	\bibfield  {author} {\bibinfo {author} {\bibfnamefont {F.}~\bibnamefont
			{Campaioli}}, \bibinfo {author} {\bibfnamefont {S.}~\bibnamefont
			{Gherardini}}, \bibinfo {author} {\bibfnamefont {J.~Q.}\ \bibnamefont
			{Quach}}, \bibinfo {author} {\bibfnamefont {M.}~\bibnamefont {Polini}}, \
		and\ \bibinfo {author} {\bibfnamefont {G.~M.}\ \bibnamefont {Andolina}},\
	}\bibinfo {title} {Colloquium: Quantum batteries},\ \href {\doibase
		10.1103/RevModPhys.96.031001} {\bibfield  {journal} {\bibinfo  {journal}
			{Rev. Mod. Phys.}\ }\textbf {\bibinfo {volume} {96}},\ \bibinfo {pages}
		{031001} (\bibinfo {year} {2024})}\BibitemShut {NoStop}%
	\bibitem [{\citenamefont {Lu}\ \emph {et~al.}(2025)\citenamefont {Lu},
		\citenamefont {Tian}, \citenamefont {L\"u},\ and\ \citenamefont
		{Shang}}]{PhysRevLett.134.180401}%
	\BibitemOpen
	\bibfield  {author} {\bibinfo {author} {\bibfnamefont {Z.-G.}\ \bibnamefont
			{Lu}}, \bibinfo {author} {\bibfnamefont {G.}~\bibnamefont {Tian}}, \bibinfo
		{author} {\bibfnamefont {X.-Y.}\ \bibnamefont {L\"u}}, \ and\ \bibinfo
		{author} {\bibfnamefont {C.}~\bibnamefont {Shang}},\ }\bibinfo {title}
	{Topological Quantum Batteries},\ \href {\doibase
		10.1103/PhysRevLett.134.180401} {\bibfield  {journal} {\bibinfo  {journal}
			{Phys. Rev. Lett.}\ }\textbf {\bibinfo {volume} {134}},\ \bibinfo {pages}
		{180401} (\bibinfo {year} {2025})}\BibitemShut {NoStop}%
	\bibitem [{\citenamefont {Arjmandi}\ \emph {et~al.}(2023)\citenamefont
		{Arjmandi}, \citenamefont {Mohammadi}, \citenamefont {Saguia}, \citenamefont
		{Sarandy},\ and\ \citenamefont {Santos}}]{PhysRevE.108.064106}%
	\BibitemOpen
	\bibfield  {author} {\bibinfo {author} {\bibfnamefont {M.~B.}\ \bibnamefont
			{Arjmandi}}, \bibinfo {author} {\bibfnamefont {H.}~\bibnamefont {Mohammadi}},
		\bibinfo {author} {\bibfnamefont {A.}~\bibnamefont {Saguia}}, \bibinfo
		{author} {\bibfnamefont {M.~S.}\ \bibnamefont {Sarandy}}, \ and\ \bibinfo
		{author} {\bibfnamefont {A.~C.}\ \bibnamefont {Santos}},\ }\bibinfo {title}
	{Localization effects in disordered quantum batteries},\ \href {\doibase
		10.1103/PhysRevE.108.064106} {\bibfield  {journal} {\bibinfo  {journal}
			{Phys. Rev. E}\ }\textbf {\bibinfo {volume} {108}},\ \bibinfo {pages}
		{064106} (\bibinfo {year} {2023})}\BibitemShut {NoStop}%
	\bibitem [{\citenamefont {Ukhtary}\ \emph {et~al.}(2023)\citenamefont
		{Ukhtary}, \citenamefont {Nugraha}, \citenamefont {Cahaya}, \citenamefont
		{Rusydi},\ and\ \citenamefont {Majidi}}]{10.1063/5.0156618}%
	\BibitemOpen
	\bibfield  {author} {\bibinfo {author} {\bibfnamefont {M.~S.}\ \bibnamefont
			{Ukhtary}}, \bibinfo {author} {\bibfnamefont {A.~R.~T.}\ \bibnamefont
			{Nugraha}}, \bibinfo {author} {\bibfnamefont {A.~B.}\ \bibnamefont {Cahaya}},
		\bibinfo {author} {\bibfnamefont {A.}~\bibnamefont {Rusydi}}, \ and\ \bibinfo
		{author} {\bibfnamefont {M.~A.}\ \bibnamefont {Majidi}},\ }\bibinfo {title}
	{High-performance Kerr quantum battery},\ \href {\doibase 10.1063/5.0156618}
	{\bibfield  {journal} {\bibinfo  {journal} {Applied Physics Letters}\
		}\textbf {\bibinfo {volume} {123}},\ \bibinfo {pages} {034001} (\bibinfo
		{year} {2023})}\BibitemShut {NoStop}%
	\bibitem [{\citenamefont {Downing}\ and\ \citenamefont
		{Ukhtary}(2023)}]{downing2023quantum}%
	\BibitemOpen
	\bibfield  {author} {\bibinfo {author} {\bibfnamefont {C.~A.}\ \bibnamefont
			{Downing}}\ and\ \bibinfo {author} {\bibfnamefont {M.~S.}\ \bibnamefont
			{Ukhtary}},\ }\bibinfo {title} {A quantum battery with quadratic driving},\
	\href@noop {} {\bibfield  {journal} {\bibinfo  {journal} {Communications
				Physics}\ }\textbf {\bibinfo {volume} {6}},\ \bibinfo {pages} {322} (\bibinfo
		{year} {2023})}\BibitemShut {NoStop}%
	\bibitem [{\citenamefont {Downing}\ and\ \citenamefont
		{Ukhtary}(2024{\natexlab{a}})}]{PhysRevA.109.052206}%
	\BibitemOpen
	\bibfield  {author} {\bibinfo {author} {\bibfnamefont {C.~A.}\ \bibnamefont
			{Downing}}\ and\ \bibinfo {author} {\bibfnamefont {M.~S.}\ \bibnamefont
			{Ukhtary}},\ }\bibinfo {title} {Hyperbolic enhancement of a quantum
		battery},\ \href {\doibase 10.1103/PhysRevA.109.052206} {\bibfield  {journal}
		{\bibinfo  {journal} {Phys. Rev. A}\ }\textbf {\bibinfo {volume} {109}},\
		\bibinfo {pages} {052206} (\bibinfo {year} {2024}{\natexlab{a}})}\BibitemShut
	{NoStop}%
	\bibitem [{\citenamefont {Downing}\ and\ \citenamefont
		{Ukhtary}(2024{\natexlab{b}})}]{Downing_2024}%
	\BibitemOpen
	\bibfield  {author} {\bibinfo {author} {\bibfnamefont {C.~A.}\ \bibnamefont
			{Downing}}\ and\ \bibinfo {author} {\bibfnamefont {M.~S.}\ \bibnamefont
			{Ukhtary}},\ }\bibinfo {title} {Energetics of a pulsed quantum battery},\
	\href {\doibase 10.1209/0295-5075/ad2e79} {\bibfield  {journal} {\bibinfo
			{journal} {Europhysics Letters}\ }\textbf {\bibinfo {volume} {146}},\
		\bibinfo {pages} {10001} (\bibinfo {year} {2024}{\natexlab{b}})}\BibitemShut
	{NoStop}%
	\bibitem [{\citenamefont {Song}\ \emph {et~al.}(2024)\citenamefont {Song},
		\citenamefont {Liu}, \citenamefont {Zhou}, \citenamefont {Yang},\ and\
		\citenamefont {An}}]{PhysRevLett.132.090401}%
	\BibitemOpen
	\bibfield  {author} {\bibinfo {author} {\bibfnamefont {W.-L.}\ \bibnamefont
			{Song}}, \bibinfo {author} {\bibfnamefont {H.-B.}\ \bibnamefont {Liu}},
		\bibinfo {author} {\bibfnamefont {B.}~\bibnamefont {Zhou}}, \bibinfo {author}
		{\bibfnamefont {W.-L.}\ \bibnamefont {Yang}}, \ and\ \bibinfo {author}
		{\bibfnamefont {J.-H.}\ \bibnamefont {An}},\ }\bibinfo {title} {Remote
		Charging and Degradation Suppression for the Quantum Battery},\ \href
	{\doibase 10.1103/PhysRevLett.132.090401} {\bibfield  {journal} {\bibinfo
			{journal} {Phys. Rev. Lett.}\ }\textbf {\bibinfo {volume} {132}},\ \bibinfo
		{pages} {090401} (\bibinfo {year} {2024})}\BibitemShut {NoStop}%
	\bibitem [{\citenamefont {Ahmadi}\ \emph {et~al.}(2024)\citenamefont {Ahmadi},
		\citenamefont {Mazurek}, \citenamefont {Horodecki},\ and\ \citenamefont
		{Barzanjeh}}]{PhysRevLett.132.210402}%
	\BibitemOpen
	\bibfield  {author} {\bibinfo {author} {\bibfnamefont {B.}~\bibnamefont
			{Ahmadi}}, \bibinfo {author} {\bibfnamefont {P.}~\bibnamefont {Mazurek}},
		\bibinfo {author} {\bibfnamefont {P.}~\bibnamefont {Horodecki}}, \ and\
		\bibinfo {author} {\bibfnamefont {S.}~\bibnamefont {Barzanjeh}},\ }\bibinfo
	{title} {Nonreciprocal Quantum Batteries},\ \href {\doibase
		10.1103/PhysRevLett.132.210402} {\bibfield  {journal} {\bibinfo  {journal}
			{Phys. Rev. Lett.}\ }\textbf {\bibinfo {volume} {132}},\ \bibinfo {pages}
		{210402} (\bibinfo {year} {2024})}\BibitemShut {NoStop}%
	\bibitem [{\citenamefont {Khan}\ \emph {et~al.}(2025)\citenamefont {Khan},
		\citenamefont {Zhang}, \citenamefont {Huang}, \citenamefont {Liu},\ and\
		\citenamefont {He}}]{67wh-1fxv}%
	\BibitemOpen
	\bibfield  {author} {\bibinfo {author} {\bibfnamefont {N.~A.}\ \bibnamefont
			{Khan}}, \bibinfo {author} {\bibfnamefont {X.}~\bibnamefont {Zhang}},
		\bibinfo {author} {\bibfnamefont {C.}~\bibnamefont {Huang}}, \bibinfo
		{author} {\bibfnamefont {Y.}~\bibnamefont {Liu}}, \ and\ \bibinfo {author}
		{\bibfnamefont {D.}~\bibnamefont {He}},\ }\bibinfo {title} {Collective
		enhancement in nonreciprocal multimode quantum batteries},\ \href {\doibase
		10.1103/67wh-1fxv} {\bibfield  {journal} {\bibinfo  {journal} {Phys. Rev. B}\
		}\textbf {\bibinfo {volume} {112}},\ \bibinfo {pages} {104318} (\bibinfo
		{year} {2025})}\BibitemShut {NoStop}%
	\bibitem [{\citenamefont {Hadipour}\ \emph {et~al.}(2025)\citenamefont
		{Hadipour}, \citenamefont {Yousefi}, \citenamefont {Mortezapour},
		\citenamefont {Miavaghi},\ and\ \citenamefont
		{Haseli}}]{hadipour2025amplified}%
	\BibitemOpen
	\bibfield  {author} {\bibinfo {author} {\bibfnamefont {M.}~\bibnamefont
			{Hadipour}}, \bibinfo {author} {\bibfnamefont {N.~N.}\ \bibnamefont
			{Yousefi}}, \bibinfo {author} {\bibfnamefont {A.}~\bibnamefont
			{Mortezapour}}, \bibinfo {author} {\bibfnamefont {A.~S.}\ \bibnamefont
			{Miavaghi}}, \ and\ \bibinfo {author} {\bibfnamefont {S.}~\bibnamefont
			{Haseli}},\ }\bibinfo {title} {Amplified quantum battery via dynamical
		modulation},\ \href@noop {} {\bibfield  {journal} {\bibinfo  {journal}
			{Scientific Reports}\ }\textbf {\bibinfo {volume} {15}},\ \bibinfo {pages}
		{14578} (\bibinfo {year} {2025})}\BibitemShut {NoStop}%
	\bibitem [{\citenamefont {Zhao}\ \emph {et~al.}(2025)\citenamefont {Zhao},
		\citenamefont {Zhao},\ and\ \citenamefont {Zhuang}}]{xqtv-qbyk}%
	\BibitemOpen
	\bibfield  {author} {\bibinfo {author} {\bibfnamefont {S.-C.}\ \bibnamefont
			{Zhao}}, \bibinfo {author} {\bibfnamefont {Z.-R.}\ \bibnamefont {Zhao}}, \
		and\ \bibinfo {author} {\bibfnamefont {N.-Y.}\ \bibnamefont {Zhuang}},\
	}\bibinfo {title} {Non-Markovian $N$-spin chain quantum battery in thermal
		charging process},\ \href {\doibase 10.1103/xqtv-qbyk} {\bibfield  {journal}
		{\bibinfo  {journal} {Phys. Rev. E}\ }\textbf {\bibinfo {volume} {112}},\
		\bibinfo {pages} {024129} (\bibinfo {year} {2025})}\BibitemShut {NoStop}%
	\bibitem [{\citenamefont {Camposeo}\ \emph {et~al.}(2025)\citenamefont
		{Camposeo}, \citenamefont {Virgili}, \citenamefont {Lombardi}, \citenamefont
		{Cerullo}, \citenamefont {Pisignano},\ and\ \citenamefont
		{Polini}}]{10.1002/adma.202415073}%
	\BibitemOpen
	\bibfield  {author} {\bibinfo {author} {\bibfnamefont {A.}~\bibnamefont
			{Camposeo}}, \bibinfo {author} {\bibfnamefont {T.}~\bibnamefont {Virgili}},
		\bibinfo {author} {\bibfnamefont {F.}~\bibnamefont {Lombardi}}, \bibinfo
		{author} {\bibfnamefont {G.}~\bibnamefont {Cerullo}}, \bibinfo {author}
		{\bibfnamefont {D.}~\bibnamefont {Pisignano}}, \ and\ \bibinfo {author}
		{\bibfnamefont {M.}~\bibnamefont {Polini}},\ }\bibinfo {title} {Quantum
		Batteries: A Materials Science Perspective},\ \href {\doibase
		https://doi.org/10.1002/adma.202415073} {\bibfield  {journal} {\bibinfo
			{journal} {Advanced Materials}\ }\textbf {\bibinfo {volume} {37}},\ \bibinfo
		{pages} {2415073} (\bibinfo {year} {2025})}\BibitemShut {NoStop}%
	\bibitem [{\citenamefont {Ferraro}\ \emph {et~al.}(2026)\citenamefont
		{Ferraro}, \citenamefont {Cavaliere}, \citenamefont {Genoni}, \citenamefont
		{Benenti},\ and\ \citenamefont {Sassetti}}]{ferraro2026opportunities}%
	\BibitemOpen
	\bibfield  {author} {\bibinfo {author} {\bibfnamefont {D.}~\bibnamefont
			{Ferraro}}, \bibinfo {author} {\bibfnamefont {F.}~\bibnamefont {Cavaliere}},
		\bibinfo {author} {\bibfnamefont {M.~G.}\ \bibnamefont {Genoni}}, \bibinfo
		{author} {\bibfnamefont {G.}~\bibnamefont {Benenti}}, \ and\ \bibinfo
		{author} {\bibfnamefont {M.}~\bibnamefont {Sassetti}},\ }\bibinfo {title}
	{Opportunities and challenges of quantum batteries},\ \href@noop {}
	{\bibfield  {journal} {\bibinfo  {journal} {Nature Reviews Physics}\ ,\
			\bibinfo {pages} {1}} (\bibinfo {year} {2026})}\BibitemShut {NoStop}%
	\bibitem [{\citenamefont {Zafar}\ and\ \citenamefont
		{Irfan}(2026)}]{10.1002/qute.202500845}%
	\BibitemOpen
	\bibfield  {author} {\bibinfo {author} {\bibfnamefont {M.~Z.}\ \bibnamefont
			{Zafar}}\ and\ \bibinfo {author} {\bibfnamefont {M.}~\bibnamefont {Irfan}},\
	}\bibinfo {title} {Loss-Induced Nonreciprocal Quantum Battery},\ \href
	{\doibase https://doi.org/10.1002/qute.202500845} {\bibfield  {journal}
		{\bibinfo  {journal} {Advanced Quantum Technologies}\ }\textbf {\bibinfo
			{volume} {9}},\ \bibinfo {pages} {e00845} (\bibinfo {year}
		{2026})}\BibitemShut {NoStop}%
	\bibitem [{\citenamefont {Monsel}\ \emph {et~al.}(2020)\citenamefont {Monsel},
		\citenamefont {Fellous-Asiani}, \citenamefont {Huard},\ and\ \citenamefont
		{Auff\`eves}}]{PhysRevLett.124.130601}%
	\BibitemOpen
	\bibfield  {author} {\bibinfo {author} {\bibfnamefont {J.}~\bibnamefont
			{Monsel}}, \bibinfo {author} {\bibfnamefont {M.}~\bibnamefont
			{Fellous-Asiani}}, \bibinfo {author} {\bibfnamefont {B.}~\bibnamefont
			{Huard}}, \ and\ \bibinfo {author} {\bibfnamefont {A.}~\bibnamefont
			{Auff\`eves}},\ }\bibinfo {title} {The Energetic Cost of Work Extraction},\
	\href {\doibase 10.1103/PhysRevLett.124.130601} {\bibfield  {journal}
		{\bibinfo  {journal} {Phys. Rev. Lett.}\ }\textbf {\bibinfo {volume} {124}},\
		\bibinfo {pages} {130601} (\bibinfo {year} {2020})}\BibitemShut {NoStop}%
	\bibitem [{\citenamefont {Yang}\ \emph {et~al.}(2023)\citenamefont {Yang},
		\citenamefont {Yang}, \citenamefont {Alimuddin}, \citenamefont {Salvia},
		\citenamefont {Fei}, \citenamefont {Zhao}, \citenamefont {Nimmrichter},\ and\
		\citenamefont {Luo}}]{PhysRevLett.131.030402}%
	\BibitemOpen
	\bibfield  {author} {\bibinfo {author} {\bibfnamefont {X.}~\bibnamefont
			{Yang}}, \bibinfo {author} {\bibfnamefont {Y.-H.}\ \bibnamefont {Yang}},
		\bibinfo {author} {\bibfnamefont {M.}~\bibnamefont {Alimuddin}}, \bibinfo
		{author} {\bibfnamefont {R.}~\bibnamefont {Salvia}}, \bibinfo {author}
		{\bibfnamefont {S.-M.}\ \bibnamefont {Fei}}, \bibinfo {author} {\bibfnamefont
			{L.-M.}\ \bibnamefont {Zhao}}, \bibinfo {author} {\bibfnamefont
			{S.}~\bibnamefont {Nimmrichter}}, \ and\ \bibinfo {author} {\bibfnamefont
			{M.-X.}\ \bibnamefont {Luo}},\ }\bibinfo {title} {Battery Capacity of
		Energy-Storing Quantum Systems},\ \href {\doibase
		10.1103/PhysRevLett.131.030402} {\bibfield  {journal} {\bibinfo  {journal}
			{Phys. Rev. Lett.}\ }\textbf {\bibinfo {volume} {131}},\ \bibinfo {pages}
		{030402} (\bibinfo {year} {2023})}\BibitemShut {NoStop}%
	\bibitem [{\citenamefont {Bhattacharyya}\ \emph {et~al.}(2024)\citenamefont
		{Bhattacharyya}, \citenamefont {Sen},\ and\ \citenamefont
		{Sen}}]{PhysRevLett.132.240401}%
	\BibitemOpen
	\bibfield  {author} {\bibinfo {author} {\bibfnamefont {A.}~\bibnamefont
			{Bhattacharyya}}, \bibinfo {author} {\bibfnamefont {K.}~\bibnamefont {Sen}},
		\ and\ \bibinfo {author} {\bibfnamefont {U.}~\bibnamefont {Sen}},\ }\bibinfo
	{title} {Noncompletely Positive Quantum Maps Enable Efficient Local Energy
		Extraction in Batteries},\ \href {\doibase 10.1103/PhysRevLett.132.240401}
	{\bibfield  {journal} {\bibinfo  {journal} {Phys. Rev. Lett.}\ }\textbf
		{\bibinfo {volume} {132}},\ \bibinfo {pages} {240401} (\bibinfo {year}
		{2024})}\BibitemShut {NoStop}%
	\bibitem [{\citenamefont {Yu}\ \emph {et~al.}(2024)\citenamefont {Yu},
		\citenamefont {Wang}, \citenamefont {Liu}, \citenamefont {Zha}, \citenamefont
		{Wu}, \citenamefont {Chen}, \citenamefont {Ye}, \citenamefont {Li},
		\citenamefont {Zhu}, \citenamefont {Guo}, \citenamefont {Qian}, \citenamefont
		{Huang}, \citenamefont {Zhao}, \citenamefont {Ying}, \citenamefont {Fan},
		\citenamefont {Wu}, \citenamefont {Su}, \citenamefont {Deng}, \citenamefont
		{Rong}, \citenamefont {Zhang}, \citenamefont {Cao}, \citenamefont {Lin},
		\citenamefont {Xu}, \citenamefont {Guo}, \citenamefont {Li}, \citenamefont
		{Liang}, \citenamefont {Wu}, \citenamefont {Huo}, \citenamefont {Lu},
		\citenamefont {Peng}, \citenamefont {Nemoto}, \citenamefont {Munro},
		\citenamefont {Zhu}, \citenamefont {Pan},\ and\ \citenamefont
		{Gong}}]{PhysRevA.109.062614}%
	\BibitemOpen
	\bibfield  {author} {\bibinfo {author} {\bibfnamefont {J.}~\bibnamefont
			{Yu}}, \bibinfo {author} {\bibfnamefont {S.}~\bibnamefont {Wang}}, \bibinfo
		{author} {\bibfnamefont {K.}~\bibnamefont {Liu}}, \bibinfo {author}
		{\bibfnamefont {C.}~\bibnamefont {Zha}}, \bibinfo {author} {\bibfnamefont
			{Y.}~\bibnamefont {Wu}}, \bibinfo {author} {\bibfnamefont {F.}~\bibnamefont
			{Chen}}, \bibinfo {author} {\bibfnamefont {Y.}~\bibnamefont {Ye}}, \bibinfo
		{author} {\bibfnamefont {S.}~\bibnamefont {Li}}, \bibinfo {author}
		{\bibfnamefont {Q.}~\bibnamefont {Zhu}}, \bibinfo {author} {\bibfnamefont
			{S.}~\bibnamefont {Guo}}, \bibinfo {author} {\bibfnamefont {H.}~\bibnamefont
			{Qian}}, \bibinfo {author} {\bibfnamefont {H.-L.}\ \bibnamefont {Huang}},
		\bibinfo {author} {\bibfnamefont {Y.}~\bibnamefont {Zhao}}, \bibinfo {author}
		{\bibfnamefont {C.}~\bibnamefont {Ying}}, \bibinfo {author} {\bibfnamefont
			{D.}~\bibnamefont {Fan}}, \bibinfo {author} {\bibfnamefont {D.}~\bibnamefont
			{Wu}}, \bibinfo {author} {\bibfnamefont {H.}~\bibnamefont {Su}}, \bibinfo
		{author} {\bibfnamefont {H.}~\bibnamefont {Deng}}, \bibinfo {author}
		{\bibfnamefont {H.}~\bibnamefont {Rong}}, \bibinfo {author} {\bibfnamefont
			{K.}~\bibnamefont {Zhang}}, \bibinfo {author} {\bibfnamefont
			{S.}~\bibnamefont {Cao}}, \bibinfo {author} {\bibfnamefont {J.}~\bibnamefont
			{Lin}}, \bibinfo {author} {\bibfnamefont {Y.}~\bibnamefont {Xu}}, \bibinfo
		{author} {\bibfnamefont {C.}~\bibnamefont {Guo}}, \bibinfo {author}
		{\bibfnamefont {N.}~\bibnamefont {Li}}, \bibinfo {author} {\bibfnamefont
			{F.}~\bibnamefont {Liang}}, \bibinfo {author} {\bibfnamefont
			{G.}~\bibnamefont {Wu}}, \bibinfo {author} {\bibfnamefont {Y.-H.}\
			\bibnamefont {Huo}}, \bibinfo {author} {\bibfnamefont {C.-Y.}\ \bibnamefont
			{Lu}}, \bibinfo {author} {\bibfnamefont {C.-Z.}\ \bibnamefont {Peng}},
		\bibinfo {author} {\bibfnamefont {K.}~\bibnamefont {Nemoto}}, \bibinfo
		{author} {\bibfnamefont {W.~J.}\ \bibnamefont {Munro}}, \bibinfo {author}
		{\bibfnamefont {X.}~\bibnamefont {Zhu}}, \bibinfo {author} {\bibfnamefont
			{J.-W.}\ \bibnamefont {Pan}}, \ and\ \bibinfo {author} {\bibfnamefont
			{M.}~\bibnamefont {Gong}},\ }\bibinfo {title} {Experimental demonstration of
		a Maxwell's demon quantum battery in a superconducting noisy
		intermediate-scale quantum processor},\ \href {\doibase
		10.1103/PhysRevA.109.062614} {\bibfield  {journal} {\bibinfo  {journal}
			{Phys. Rev. A}\ }\textbf {\bibinfo {volume} {109}},\ \bibinfo {pages}
		{062614} (\bibinfo {year} {2024})}\BibitemShut {NoStop}%
	\bibitem [{\citenamefont {Juli\`a-Farr\'e}\ \emph {et~al.}(2020)\citenamefont
		{Juli\`a-Farr\'e}, \citenamefont {Salamon}, \citenamefont {Riera},
		\citenamefont {Bera},\ and\ \citenamefont
		{Lewenstein}}]{PhysRevResearch.2.023113}%
	\BibitemOpen
	\bibfield  {author} {\bibinfo {author} {\bibfnamefont {S.}~\bibnamefont
			{Juli\`a-Farr\'e}}, \bibinfo {author} {\bibfnamefont {T.}~\bibnamefont
			{Salamon}}, \bibinfo {author} {\bibfnamefont {A.}~\bibnamefont {Riera}},
		\bibinfo {author} {\bibfnamefont {M.~N.}\ \bibnamefont {Bera}}, \ and\
		\bibinfo {author} {\bibfnamefont {M.}~\bibnamefont {Lewenstein}},\ }\bibinfo
	{title} {Bounds on the capacity and power of quantum batteries},\ \href
	{\doibase 10.1103/PhysRevResearch.2.023113} {\bibfield  {journal} {\bibinfo
			{journal} {Phys. Rev. Res.}\ }\textbf {\bibinfo {volume} {2}},\ \bibinfo
		{pages} {023113} (\bibinfo {year} {2020})}\BibitemShut {NoStop}%
	\bibitem [{\citenamefont {Quach}\ and\ \citenamefont
		{Munro}(2020)}]{PhysRevApplied.14.024092}%
	\BibitemOpen
	\bibfield  {author} {\bibinfo {author} {\bibfnamefont {J.~Q.}\ \bibnamefont
			{Quach}}\ and\ \bibinfo {author} {\bibfnamefont {W.~J.}\ \bibnamefont
			{Munro}},\ }\bibinfo {title} {Using Dark States to Charge and Stabilize Open
		Quantum Batteries},\ \href {\doibase 10.1103/PhysRevApplied.14.024092}
	{\bibfield  {journal} {\bibinfo  {journal} {Phys. Rev. Appl.}\ }\textbf
		{\bibinfo {volume} {14}},\ \bibinfo {pages} {024092} (\bibinfo {year}
		{2020})}\BibitemShut {NoStop}%
	\bibitem [{\citenamefont {Santos}\ \emph {et~al.}(2020)\citenamefont {Santos},
		\citenamefont {Saguia},\ and\ \citenamefont {Sarandy}}]{PhysRevE.101.062114}%
	\BibitemOpen
	\bibfield  {author} {\bibinfo {author} {\bibfnamefont {A.~C.}\ \bibnamefont
			{Santos}}, \bibinfo {author} {\bibfnamefont {A.}~\bibnamefont {Saguia}}, \
		and\ \bibinfo {author} {\bibfnamefont {M.~S.}\ \bibnamefont {Sarandy}},\
	}\bibinfo {title} {Stable and charge-switchable quantum batteries},\ \href
	{\doibase 10.1103/PhysRevE.101.062114} {\bibfield  {journal} {\bibinfo
			{journal} {Phys. Rev. E}\ }\textbf {\bibinfo {volume} {101}},\ \bibinfo
		{pages} {062114} (\bibinfo {year} {2020})}\BibitemShut {NoStop}%
	\bibitem [{\citenamefont {Sun}\ \emph {et~al.}(2025)\citenamefont {Sun},
		\citenamefont {Wang}, \citenamefont {Yan}, \citenamefont {Zhang},
		\citenamefont {Man},\ and\ \citenamefont {Cai}}]{p93y-jflt}%
	\BibitemOpen
	\bibfield  {author} {\bibinfo {author} {\bibfnamefont {C.-Z.}\ \bibnamefont
			{Sun}}, \bibinfo {author} {\bibfnamefont {Z.-K.}\ \bibnamefont {Wang}},
		\bibinfo {author} {\bibfnamefont {W.-B.}\ \bibnamefont {Yan}}, \bibinfo
		{author} {\bibfnamefont {Y.-J.}\ \bibnamefont {Zhang}}, \bibinfo {author}
		{\bibfnamefont {Z.-X.}\ \bibnamefont {Man}}, \ and\ \bibinfo {author}
		{\bibfnamefont {Q.-Y.}\ \bibnamefont {Cai}},\ }\bibinfo {title}
	{Nonreciprocal charging in a quantum battery via a mediator},\ \href
	{\doibase 10.1103/p93y-jflt} {\bibfield  {journal} {\bibinfo  {journal}
			{Phys. Rev. A}\ }\textbf {\bibinfo {volume} {112}},\ \bibinfo {pages}
		{012429} (\bibinfo {year} {2025})}\BibitemShut {NoStop}%
	\bibitem [{\citenamefont {Lai}\ \emph {et~al.}(2024)\citenamefont {Lai},
		\citenamefont {Lin}, \citenamefont {Huang}, \citenamefont {Jan},\ and\
		\citenamefont {Chen}}]{PhysRevResearch.6.023136}%
	\BibitemOpen
	\bibfield  {author} {\bibinfo {author} {\bibfnamefont {P.-R.}\ \bibnamefont
			{Lai}}, \bibinfo {author} {\bibfnamefont {J.-D.}\ \bibnamefont {Lin}},
		\bibinfo {author} {\bibfnamefont {Y.-T.}\ \bibnamefont {Huang}}, \bibinfo
		{author} {\bibfnamefont {H.-C.}\ \bibnamefont {Jan}}, \ and\ \bibinfo
		{author} {\bibfnamefont {Y.-N.}\ \bibnamefont {Chen}},\ }\bibinfo {title}
	{Quick charging of a quantum battery with superposed trajectories},\ \href
	{\doibase 10.1103/PhysRevResearch.6.023136} {\bibfield  {journal} {\bibinfo
			{journal} {Phys. Rev. Res.}\ }\textbf {\bibinfo {volume} {6}},\ \bibinfo
		{pages} {023136} (\bibinfo {year} {2024})}\BibitemShut {NoStop}%
	\bibitem [{\citenamefont {Zhang}\ \emph {et~al.}(2024)\citenamefont {Zhang},
		\citenamefont {Ma}, \citenamefont {Jiang}, \citenamefont {Yu}, \citenamefont
		{Jin},\ and\ \citenamefont {Chen}}]{PhysRevA.110.032211}%
	\BibitemOpen
	\bibfield  {author} {\bibinfo {author} {\bibfnamefont {D.}~\bibnamefont
			{Zhang}}, \bibinfo {author} {\bibfnamefont {S.}~\bibnamefont {Ma}}, \bibinfo
		{author} {\bibfnamefont {Y.}~\bibnamefont {Jiang}}, \bibinfo {author}
		{\bibfnamefont {Y.}~\bibnamefont {Yu}}, \bibinfo {author} {\bibfnamefont
			{G.}~\bibnamefont {Jin}}, \ and\ \bibinfo {author} {\bibfnamefont
			{A.}~\bibnamefont {Chen}},\ }\bibinfo {title} {Quantum battery with
		interactive atomic collective charging},\ \href {\doibase
		10.1103/PhysRevA.110.032211} {\bibfield  {journal} {\bibinfo  {journal}
			{Phys. Rev. A}\ }\textbf {\bibinfo {volume} {110}},\ \bibinfo {pages}
		{032211} (\bibinfo {year} {2024})}\BibitemShut {NoStop}%
	\bibitem [{\citenamefont {Shastri}\ \emph {et~al.}(2025)\citenamefont
		{Shastri}, \citenamefont {Jiang}, \citenamefont {Xu}, \citenamefont
		{Prasanna~Venkatesh},\ and\ \citenamefont {Watanabe}}]{shastri2025dephasing}%
	\BibitemOpen
	\bibfield  {author} {\bibinfo {author} {\bibfnamefont {R.}~\bibnamefont
			{Shastri}}, \bibinfo {author} {\bibfnamefont {C.}~\bibnamefont {Jiang}},
		\bibinfo {author} {\bibfnamefont {G.-H.}\ \bibnamefont {Xu}}, \bibinfo
		{author} {\bibfnamefont {B.}~\bibnamefont {Prasanna~Venkatesh}}, \ and\
		\bibinfo {author} {\bibfnamefont {G.}~\bibnamefont {Watanabe}},\ }\bibinfo
	{title} {Dephasing enabled fast charging of quantum batteries},\ \href@noop
	{} {\bibfield  {journal} {\bibinfo  {journal} {npj Quantum Information}\
		}\textbf {\bibinfo {volume} {11}},\ \bibinfo {pages} {9} (\bibinfo {year}
		{2025})}\BibitemShut {NoStop}%
	\bibitem [{\citenamefont {Campaioli}\ \emph {et~al.}(2017)\citenamefont
		{Campaioli}, \citenamefont {Pollock}, \citenamefont {Binder}, \citenamefont
		{C\'eleri}, \citenamefont {Goold}, \citenamefont {Vinjanampathy},\ and\
		\citenamefont {Modi}}]{PhysRevLett.118.150601}%
	\BibitemOpen
	\bibfield  {author} {\bibinfo {author} {\bibfnamefont {F.}~\bibnamefont
			{Campaioli}}, \bibinfo {author} {\bibfnamefont {F.~A.}\ \bibnamefont
			{Pollock}}, \bibinfo {author} {\bibfnamefont {F.~C.}\ \bibnamefont {Binder}},
		\bibinfo {author} {\bibfnamefont {L.}~\bibnamefont {C\'eleri}}, \bibinfo
		{author} {\bibfnamefont {J.}~\bibnamefont {Goold}}, \bibinfo {author}
		{\bibfnamefont {S.}~\bibnamefont {Vinjanampathy}}, \ and\ \bibinfo {author}
		{\bibfnamefont {K.}~\bibnamefont {Modi}},\ }\bibinfo {title} {Enhancing the
		Charging Power of Quantum Batteries},\ \href {\doibase
		10.1103/PhysRevLett.118.150601} {\bibfield  {journal} {\bibinfo  {journal}
			{Phys. Rev. Lett.}\ }\textbf {\bibinfo {volume} {118}},\ \bibinfo {pages}
		{150601} (\bibinfo {year} {2017})}\BibitemShut {NoStop}%
	\bibitem [{\citenamefont {Seah}\ \emph {et~al.}(2021)\citenamefont {Seah},
		\citenamefont {Perarnau-Llobet}, \citenamefont {Haack}, \citenamefont
		{Brunner},\ and\ \citenamefont {Nimmrichter}}]{PhysRevLett.127.100601}%
	\BibitemOpen
	\bibfield  {author} {\bibinfo {author} {\bibfnamefont {S.}~\bibnamefont
			{Seah}}, \bibinfo {author} {\bibfnamefont {M.}~\bibnamefont
			{Perarnau-Llobet}}, \bibinfo {author} {\bibfnamefont {G.}~\bibnamefont
			{Haack}}, \bibinfo {author} {\bibfnamefont {N.}~\bibnamefont {Brunner}}, \
		and\ \bibinfo {author} {\bibfnamefont {S.}~\bibnamefont {Nimmrichter}},\
	}\bibinfo {title} {Quantum Speed-Up in Collisional Battery Charging},\ \href
	{\doibase 10.1103/PhysRevLett.127.100601} {\bibfield  {journal} {\bibinfo
			{journal} {Phys. Rev. Lett.}\ }\textbf {\bibinfo {volume} {127}},\ \bibinfo
		{pages} {100601} (\bibinfo {year} {2021})}\BibitemShut {NoStop}%
	\bibitem [{\citenamefont {Gyhm}\ \emph {et~al.}(2022)\citenamefont {Gyhm},
		\citenamefont {\ifmmode~\check{S}\else \v{S}\fi{}afr\'anek},\ and\
		\citenamefont {Rosa}}]{PhysRevLett.128.140501}%
	\BibitemOpen
	\bibfield  {author} {\bibinfo {author} {\bibfnamefont {J.-Y.}\ \bibnamefont
			{Gyhm}}, \bibinfo {author} {\bibfnamefont {D.}~\bibnamefont
			{\ifmmode~\check{S}\else \v{S}\fi{}afr\'anek}}, \ and\ \bibinfo {author}
		{\bibfnamefont {D.}~\bibnamefont {Rosa}},\ }\bibinfo {title} {Quantum
		Charging Advantage Cannot Be Extensive without Global Operations},\ \href
	{\doibase 10.1103/PhysRevLett.128.140501} {\bibfield  {journal} {\bibinfo
			{journal} {Phys. Rev. Lett.}\ }\textbf {\bibinfo {volume} {128}},\ \bibinfo
		{pages} {140501} (\bibinfo {year} {2022})}\BibitemShut {NoStop}%
	\bibitem [{\citenamefont {Ghosh}\ \emph {et~al.}(2020)\citenamefont {Ghosh},
		\citenamefont {Chanda},\ and\ \citenamefont {Sen(De)}}]{PhysRevA.101.032115}%
	\BibitemOpen
	\bibfield  {author} {\bibinfo {author} {\bibfnamefont {S.}~\bibnamefont
			{Ghosh}}, \bibinfo {author} {\bibfnamefont {T.}~\bibnamefont {Chanda}}, \
		and\ \bibinfo {author} {\bibfnamefont {A.}~\bibnamefont {Sen(De)}},\
	}\bibinfo {title} {Enhancement in the performance of a quantum battery by
		ordered and disordered interactions},\ \href {\doibase
		10.1103/PhysRevA.101.032115} {\bibfield  {journal} {\bibinfo  {journal}
			{Phys. Rev. A}\ }\textbf {\bibinfo {volume} {101}},\ \bibinfo {pages}
		{032115} (\bibinfo {year} {2020})}\BibitemShut {NoStop}%
	\bibitem [{\citenamefont {Crescente}\ \emph {et~al.}(2020)\citenamefont
		{Crescente}, \citenamefont {Carrega}, \citenamefont {Sassetti},\ and\
		\citenamefont {Ferraro}}]{PhysRevB.102.245407}%
	\BibitemOpen
	\bibfield  {author} {\bibinfo {author} {\bibfnamefont {A.}~\bibnamefont
			{Crescente}}, \bibinfo {author} {\bibfnamefont {M.}~\bibnamefont {Carrega}},
		\bibinfo {author} {\bibfnamefont {M.}~\bibnamefont {Sassetti}}, \ and\
		\bibinfo {author} {\bibfnamefont {D.}~\bibnamefont {Ferraro}},\ }\bibinfo
	{title} {Ultrafast charging in a two-photon Dicke quantum battery},\ \href
	{\doibase 10.1103/PhysRevB.102.245407} {\bibfield  {journal} {\bibinfo
			{journal} {Phys. Rev. B}\ }\textbf {\bibinfo {volume} {102}},\ \bibinfo
		{pages} {245407} (\bibinfo {year} {2020})}\BibitemShut {NoStop}%
	\bibitem [{\citenamefont {Huangfu}\ and\ \citenamefont
		{Jing}(2021)}]{PhysRevE.104.024129}%
	\BibitemOpen
	\bibfield  {author} {\bibinfo {author} {\bibfnamefont {Y.}~\bibnamefont
			{Huangfu}}\ and\ \bibinfo {author} {\bibfnamefont {J.}~\bibnamefont {Jing}},\
	}\bibinfo {title} {High-capacity and high-power collective charging with spin
		chargers},\ \href {\doibase 10.1103/PhysRevE.104.024129} {\bibfield
		{journal} {\bibinfo  {journal} {Phys. Rev. E}\ }\textbf {\bibinfo {volume}
			{104}},\ \bibinfo {pages} {024129} (\bibinfo {year} {2021})}\BibitemShut
	{NoStop}%
	\bibitem [{\citenamefont {Ghosh}\ \emph {et~al.}(2021)\citenamefont {Ghosh},
		\citenamefont {Chanda}, \citenamefont {Mal},\ and\ \citenamefont
		{Sen(De)}}]{PhysRevA.104.032207}%
	\BibitemOpen
	\bibfield  {author} {\bibinfo {author} {\bibfnamefont {S.}~\bibnamefont
			{Ghosh}}, \bibinfo {author} {\bibfnamefont {T.}~\bibnamefont {Chanda}},
		\bibinfo {author} {\bibfnamefont {S.}~\bibnamefont {Mal}}, \ and\ \bibinfo
		{author} {\bibfnamefont {A.}~\bibnamefont {Sen(De)}},\ }\bibinfo {title}
	{Fast charging of a quantum battery assisted by noise},\ \href {\doibase
		10.1103/PhysRevA.104.032207} {\bibfield  {journal} {\bibinfo  {journal}
			{Phys. Rev. A}\ }\textbf {\bibinfo {volume} {104}},\ \bibinfo {pages}
		{032207} (\bibinfo {year} {2021})}\BibitemShut {NoStop}%
	\bibitem [{\citenamefont {Binder}\ \emph {et~al.}(2015)\citenamefont {Binder},
		\citenamefont {Vinjanampathy}, \citenamefont {Modi},\ and\ \citenamefont
		{Goold}}]{Binder_2015}%
	\BibitemOpen
	\bibfield  {author} {\bibinfo {author} {\bibfnamefont {F.~C.}\ \bibnamefont
			{Binder}}, \bibinfo {author} {\bibfnamefont {S.}~\bibnamefont
			{Vinjanampathy}}, \bibinfo {author} {\bibfnamefont {K.}~\bibnamefont {Modi}},
		\ and\ \bibinfo {author} {\bibfnamefont {J.}~\bibnamefont {Goold}},\
	}\bibinfo {title} {Quantacell: powerful charging of quantum batteries},\
	\href {\doibase 10.1088/1367-2630/17/7/075015} {\bibfield  {journal}
		{\bibinfo  {journal} {New Journal of Physics}\ }\textbf {\bibinfo {volume}
			{17}},\ \bibinfo {pages} {075015} (\bibinfo {year} {2015})}\BibitemShut
	{NoStop}%
	\bibitem [{\citenamefont {Huang}\ \emph {et~al.}(2026)\citenamefont {Huang},
		\citenamefont {Zhang}, \citenamefont {Wang}, \citenamefont {Zhao},
		\citenamefont {Yu}, \citenamefont {Jin},\ and\ \citenamefont
		{Chen}}]{qhz8-mvfb}%
	\BibitemOpen
	\bibfield  {author} {\bibinfo {author} {\bibfnamefont {Z.}~\bibnamefont
			{Huang}}, \bibinfo {author} {\bibfnamefont {D.}~\bibnamefont {Zhang}},
		\bibinfo {author} {\bibfnamefont {Z.}~\bibnamefont {Wang}}, \bibinfo {author}
		{\bibfnamefont {Y.}~\bibnamefont {Zhao}}, \bibinfo {author} {\bibfnamefont
			{Y.}~\bibnamefont {Yu}}, \bibinfo {author} {\bibfnamefont {G.}~\bibnamefont
			{Jin}}, \ and\ \bibinfo {author} {\bibfnamefont {A.}~\bibnamefont {Chen}},\
	}\bibinfo {title} {Multilevel quantum batteries: Large battery capacity and
		high charging speed},\ \href {\doibase 10.1103/qhz8-mvfb} {\bibfield
		{journal} {\bibinfo  {journal} {Phys. Rev. A}\ }\textbf {\bibinfo {volume}
			{113}},\ \bibinfo {pages} {042616} (\bibinfo {year} {2026})}\BibitemShut
	{NoStop}%
	\bibitem [{\citenamefont {Zhu}\ \emph {et~al.}(2023)\citenamefont {Zhu},
		\citenamefont {Chen}, \citenamefont {Hasegawa},\ and\ \citenamefont
		{Xue}}]{PhysRevLett.131.240401}%
	\BibitemOpen
	\bibfield  {author} {\bibinfo {author} {\bibfnamefont {G.}~\bibnamefont
			{Zhu}}, \bibinfo {author} {\bibfnamefont {Y.}~\bibnamefont {Chen}}, \bibinfo
		{author} {\bibfnamefont {Y.}~\bibnamefont {Hasegawa}}, \ and\ \bibinfo
		{author} {\bibfnamefont {P.}~\bibnamefont {Xue}},\ }\bibinfo {title}
	{Charging Quantum Batteries via Indefinite Causal Order: Theory and
		Experiment},\ \href {\doibase 10.1103/PhysRevLett.131.240401} {\bibfield
		{journal} {\bibinfo  {journal} {Phys. Rev. Lett.}\ }\textbf {\bibinfo
			{volume} {131}},\ \bibinfo {pages} {240401} (\bibinfo {year}
		{2023})}\BibitemShut {NoStop}%
	\bibitem [{\citenamefont {Simon}\ \emph {et~al.}(2025)\citenamefont {Simon},
		\citenamefont {Anders},\ and\ \citenamefont
		{Hovhannisyan}}]{PhysRevLett.134.010408}%
	\BibitemOpen
	\bibfield  {author} {\bibinfo {author} {\bibfnamefont {R.~P.~A.}\
			\bibnamefont {Simon}}, \bibinfo {author} {\bibfnamefont {J.}~\bibnamefont
			{Anders}}, \ and\ \bibinfo {author} {\bibfnamefont {K.~V.}\ \bibnamefont
			{Hovhannisyan}},\ }\bibinfo {title} {Correlations Enable Lossless Ergotropy
		Transport},\ \href {\doibase 10.1103/PhysRevLett.134.010408} {\bibfield
		{journal} {\bibinfo  {journal} {Phys. Rev. Lett.}\ }\textbf {\bibinfo
			{volume} {134}},\ \bibinfo {pages} {010408} (\bibinfo {year}
		{2025})}\BibitemShut {NoStop}%
	\bibitem [{\citenamefont {Chitambar}\ and\ \citenamefont
		{Gour}(2019)}]{RevModPhys.91.025001}%
	\BibitemOpen
	\bibfield  {author} {\bibinfo {author} {\bibfnamefont {E.}~\bibnamefont
			{Chitambar}}\ and\ \bibinfo {author} {\bibfnamefont {G.}~\bibnamefont
			{Gour}},\ }\bibinfo {title} {Quantum resource theories},\ \href {\doibase
		10.1103/RevModPhys.91.025001} {\bibfield  {journal} {\bibinfo  {journal}
			{Rev. Mod. Phys.}\ }\textbf {\bibinfo {volume} {91}},\ \bibinfo {pages}
		{025001} (\bibinfo {year} {2019})}\BibitemShut {NoStop}%
	\bibitem [{\citenamefont {Kamin}\ \emph
		{et~al.}(2020{\natexlab{a}})\citenamefont {Kamin}, \citenamefont {Tabesh},
		\citenamefont {Salimi},\ and\ \citenamefont {Santos}}]{PhysRevE.102.052109}%
	\BibitemOpen
	\bibfield  {author} {\bibinfo {author} {\bibfnamefont {F.~H.}\ \bibnamefont
			{Kamin}}, \bibinfo {author} {\bibfnamefont {F.~T.}\ \bibnamefont {Tabesh}},
		\bibinfo {author} {\bibfnamefont {S.}~\bibnamefont {Salimi}}, \ and\ \bibinfo
		{author} {\bibfnamefont {A.~C.}\ \bibnamefont {Santos}},\ }\bibinfo {title}
	{Entanglement, coherence, and charging process of quantum batteries},\ \href
	{\doibase 10.1103/PhysRevE.102.052109} {\bibfield  {journal} {\bibinfo
			{journal} {Phys. Rev. E}\ }\textbf {\bibinfo {volume} {102}},\ \bibinfo
		{pages} {052109} (\bibinfo {year} {2020}{\natexlab{a}})}\BibitemShut
	{NoStop}%
	\bibitem [{\citenamefont {Arjmandi}\ \emph
		{et~al.}(2022{\natexlab{a}})\citenamefont {Arjmandi}, \citenamefont {Shokri},
		\citenamefont {Faizi},\ and\ \citenamefont
		{Mohammadi}}]{PhysRevA.106.062609}%
	\BibitemOpen
	\bibfield  {author} {\bibinfo {author} {\bibfnamefont {M.~B.}\ \bibnamefont
			{Arjmandi}}, \bibinfo {author} {\bibfnamefont {A.}~\bibnamefont {Shokri}},
		\bibinfo {author} {\bibfnamefont {E.}~\bibnamefont {Faizi}}, \ and\ \bibinfo
		{author} {\bibfnamefont {H.}~\bibnamefont {Mohammadi}},\ }\bibinfo {title}
	{Performance of quantum batteries with correlated and uncorrelated
		chargers},\ \href {\doibase 10.1103/PhysRevA.106.062609} {\bibfield
		{journal} {\bibinfo  {journal} {Phys. Rev. A}\ }\textbf {\bibinfo {volume}
			{106}},\ \bibinfo {pages} {062609} (\bibinfo {year}
		{2022}{\natexlab{a}})}\BibitemShut {NoStop}%
	\bibitem [{\citenamefont {Shi}\ \emph {et~al.}(2022)\citenamefont {Shi},
		\citenamefont {Ding}, \citenamefont {Wan}, \citenamefont {Wang},\ and\
		\citenamefont {Yang}}]{PhysRevLett.129.130602}%
	\BibitemOpen
	\bibfield  {author} {\bibinfo {author} {\bibfnamefont {H.-L.}\ \bibnamefont
			{Shi}}, \bibinfo {author} {\bibfnamefont {S.}~\bibnamefont {Ding}}, \bibinfo
		{author} {\bibfnamefont {Q.-K.}\ \bibnamefont {Wan}}, \bibinfo {author}
		{\bibfnamefont {X.-H.}\ \bibnamefont {Wang}}, \ and\ \bibinfo {author}
		{\bibfnamefont {W.-L.}\ \bibnamefont {Yang}},\ }\bibinfo {title}
	{Entanglement, Coherence, and Extractable Work in Quantum Batteries},\ \href
	{\doibase 10.1103/PhysRevLett.129.130602} {\bibfield  {journal} {\bibinfo
			{journal} {Phys. Rev. Lett.}\ }\textbf {\bibinfo {volume} {129}},\ \bibinfo
		{pages} {130602} (\bibinfo {year} {2022})}\BibitemShut {NoStop}%
	\bibitem [{\citenamefont {Garc\'{\i}a-Pintos}\ \emph
		{et~al.}(2020)\citenamefont {Garc\'{\i}a-Pintos}, \citenamefont {Hamma},\
		and\ \citenamefont {del Campo}}]{PhysRevLett.125.040601}%
	\BibitemOpen
	\bibfield  {author} {\bibinfo {author} {\bibfnamefont {L.~P.}\ \bibnamefont
			{Garc\'{\i}a-Pintos}}, \bibinfo {author} {\bibfnamefont {A.}~\bibnamefont
			{Hamma}}, \ and\ \bibinfo {author} {\bibfnamefont {A.}~\bibnamefont {del
				Campo}},\ }\bibinfo {title} {Fluctuations in Extractable Work Bound the
		Charging Power of Quantum Batteries},\ \href {\doibase
		10.1103/PhysRevLett.125.040601} {\bibfield  {journal} {\bibinfo  {journal}
			{Phys. Rev. Lett.}\ }\textbf {\bibinfo {volume} {125}},\ \bibinfo {pages}
		{040601} (\bibinfo {year} {2020})}\BibitemShut {NoStop}%
	\bibitem [{\citenamefont {Streltsov}\ \emph {et~al.}(2017)\citenamefont
		{Streltsov}, \citenamefont {Adesso},\ and\ \citenamefont
		{Plenio}}]{RevModPhys.89.041003}%
	\BibitemOpen
	\bibfield  {author} {\bibinfo {author} {\bibfnamefont {A.}~\bibnamefont
			{Streltsov}}, \bibinfo {author} {\bibfnamefont {G.}~\bibnamefont {Adesso}}, \
		and\ \bibinfo {author} {\bibfnamefont {M.~B.}\ \bibnamefont {Plenio}},\
	}\bibinfo {title} {Colloquium: Quantum coherence as a resource},\ \href
	{\doibase 10.1103/RevModPhys.89.041003} {\bibfield  {journal} {\bibinfo
			{journal} {Rev. Mod. Phys.}\ }\textbf {\bibinfo {volume} {89}},\ \bibinfo
		{pages} {041003} (\bibinfo {year} {2017})}\BibitemShut {NoStop}%
	\bibitem [{\citenamefont {Maillette~de Buy~Wenniger}\ \emph
		{et~al.}(2023)\citenamefont {Maillette~de Buy~Wenniger}, \citenamefont
		{Thomas}, \citenamefont {Maffei}, \citenamefont {Wein}, \citenamefont {Pont},
		\citenamefont {Belabas}, \citenamefont {Prasad}, \citenamefont {Harouri},
		\citenamefont {Lema\^{\i}tre}, \citenamefont {Sagnes}, \citenamefont
		{Somaschi}, \citenamefont {Auff\`eves},\ and\ \citenamefont
		{Senellart}}]{PhysRevLett.131.260401}%
	\BibitemOpen
	\bibfield  {author} {\bibinfo {author} {\bibfnamefont {I.}~\bibnamefont
			{Maillette~de Buy~Wenniger}}, \bibinfo {author} {\bibfnamefont {S.~E.}\
			\bibnamefont {Thomas}}, \bibinfo {author} {\bibfnamefont {M.}~\bibnamefont
			{Maffei}}, \bibinfo {author} {\bibfnamefont {S.~C.}\ \bibnamefont {Wein}},
		\bibinfo {author} {\bibfnamefont {M.}~\bibnamefont {Pont}}, \bibinfo {author}
		{\bibfnamefont {N.}~\bibnamefont {Belabas}}, \bibinfo {author} {\bibfnamefont
			{S.}~\bibnamefont {Prasad}}, \bibinfo {author} {\bibfnamefont
			{A.}~\bibnamefont {Harouri}}, \bibinfo {author} {\bibfnamefont
			{A.}~\bibnamefont {Lema\^{\i}tre}}, \bibinfo {author} {\bibfnamefont
			{I.}~\bibnamefont {Sagnes}}, \bibinfo {author} {\bibfnamefont
			{N.}~\bibnamefont {Somaschi}}, \bibinfo {author} {\bibfnamefont
			{A.}~\bibnamefont {Auff\`eves}}, \ and\ \bibinfo {author} {\bibfnamefont
			{P.}~\bibnamefont {Senellart}},\ }\bibinfo {title} {Experimental Analysis of
		Energy Transfers between a Quantum Emitter and Light Fields},\ \href
	{\doibase 10.1103/PhysRevLett.131.260401} {\bibfield  {journal} {\bibinfo
			{journal} {Phys. Rev. Lett.}\ }\textbf {\bibinfo {volume} {131}},\ \bibinfo
		{pages} {260401} (\bibinfo {year} {2023})}\BibitemShut {NoStop}%
	\bibitem [{\citenamefont {Centrone}\ \emph {et~al.}(2023)\citenamefont
		{Centrone}, \citenamefont {Mancino},\ and\ \citenamefont
		{Paternostro}}]{PhysRevA.108.052213}%
	\BibitemOpen
	\bibfield  {author} {\bibinfo {author} {\bibfnamefont {F.}~\bibnamefont
			{Centrone}}, \bibinfo {author} {\bibfnamefont {L.}~\bibnamefont {Mancino}}, \
		and\ \bibinfo {author} {\bibfnamefont {M.}~\bibnamefont {Paternostro}},\
	}\bibinfo {title} {Charging batteries with quantum squeezing},\ \href
	{\doibase 10.1103/PhysRevA.108.052213} {\bibfield  {journal} {\bibinfo
			{journal} {Phys. Rev. A}\ }\textbf {\bibinfo {volume} {108}},\ \bibinfo
		{pages} {052213} (\bibinfo {year} {2023})}\BibitemShut {NoStop}%
	\bibitem [{\citenamefont {Francica}\ \emph {et~al.}(2020)\citenamefont
		{Francica}, \citenamefont {Binder}, \citenamefont {Guarnieri}, \citenamefont
		{Mitchison}, \citenamefont {Goold},\ and\ \citenamefont
		{Plastina}}]{PhysRevLett.125.180603}%
	\BibitemOpen
	\bibfield  {author} {\bibinfo {author} {\bibfnamefont {G.}~\bibnamefont
			{Francica}}, \bibinfo {author} {\bibfnamefont {F.~C.}\ \bibnamefont
			{Binder}}, \bibinfo {author} {\bibfnamefont {G.}~\bibnamefont {Guarnieri}},
		\bibinfo {author} {\bibfnamefont {M.~T.}\ \bibnamefont {Mitchison}}, \bibinfo
		{author} {\bibfnamefont {J.}~\bibnamefont {Goold}}, \ and\ \bibinfo {author}
		{\bibfnamefont {F.}~\bibnamefont {Plastina}},\ }\bibinfo {title} {Quantum
		Coherence and Ergotropy},\ \href {\doibase 10.1103/PhysRevLett.125.180603}
	{\bibfield  {journal} {\bibinfo  {journal} {Phys. Rev. Lett.}\ }\textbf
		{\bibinfo {volume} {125}},\ \bibinfo {pages} {180603} (\bibinfo {year}
		{2020})}\BibitemShut {NoStop}%
	\bibitem [{\citenamefont {Lostaglio}\ \emph {et~al.}(2015)\citenamefont
		{Lostaglio}, \citenamefont {Jennings},\ and\ \citenamefont
		{Rudolph}}]{lostaglio2015description}%
	\BibitemOpen
	\bibfield  {author} {\bibinfo {author} {\bibfnamefont {M.}~\bibnamefont
			{Lostaglio}}, \bibinfo {author} {\bibfnamefont {D.}~\bibnamefont {Jennings}},
		\ and\ \bibinfo {author} {\bibfnamefont {T.}~\bibnamefont {Rudolph}},\
	}\bibinfo {title} {Description of quantum coherence in thermodynamic
		processes requires constraints beyond free energy},\ \href@noop {} {\bibfield
		{journal} {\bibinfo  {journal} {Nature communications}\ }\textbf {\bibinfo
			{volume} {6}},\ \bibinfo {pages} {6383} (\bibinfo {year} {2015})}\BibitemShut
	{NoStop}%
	\bibitem [{\citenamefont {Uzdin}\ \emph {et~al.}(2015)\citenamefont {Uzdin},
		\citenamefont {Levy},\ and\ \citenamefont {Kosloff}}]{PhysRevX.5.031044}%
	\BibitemOpen
	\bibfield  {author} {\bibinfo {author} {\bibfnamefont {R.}~\bibnamefont
			{Uzdin}}, \bibinfo {author} {\bibfnamefont {A.}~\bibnamefont {Levy}}, \ and\
		\bibinfo {author} {\bibfnamefont {R.}~\bibnamefont {Kosloff}},\ }\bibinfo
	{title} {Equivalence of Quantum Heat Machines, and Quantum-Thermodynamic
		Signatures},\ \href {\doibase 10.1103/PhysRevX.5.031044} {\bibfield
		{journal} {\bibinfo  {journal} {Phys. Rev. X}\ }\textbf {\bibinfo {volume}
			{5}},\ \bibinfo {pages} {031044} (\bibinfo {year} {2015})}\BibitemShut
	{NoStop}%
	\bibitem [{\citenamefont {Korzekwa}\ \emph {et~al.}(2016)\citenamefont
		{Korzekwa}, \citenamefont {Lostaglio}, \citenamefont {Oppenheim},\ and\
		\citenamefont {Jennings}}]{Korzekwa_2016}%
	\BibitemOpen
	\bibfield  {author} {\bibinfo {author} {\bibfnamefont {K.}~\bibnamefont
			{Korzekwa}}, \bibinfo {author} {\bibfnamefont {M.}~\bibnamefont {Lostaglio}},
		\bibinfo {author} {\bibfnamefont {J.}~\bibnamefont {Oppenheim}}, \ and\
		\bibinfo {author} {\bibfnamefont {D.}~\bibnamefont {Jennings}},\ }\bibinfo
	{title} {The extraction of work from quantum coherence},\ \href {\doibase
		10.1088/1367-2630/18/2/023045} {\bibfield  {journal} {\bibinfo  {journal}
			{New Journal of Physics}\ }\textbf {\bibinfo {volume} {18}},\ \bibinfo
		{pages} {023045} (\bibinfo {year} {2016})}\BibitemShut {NoStop}%
	\bibitem [{\citenamefont {\ifmmode~\mbox{\c{C}}\else
			\c{C}\fi{}akmak}(2020)}]{PhysRevE.102.042111}%
	\BibitemOpen
	\bibfield  {author} {\bibinfo {author} {\bibfnamefont {B.~i. e. i. f. m.~c.}\
			\bibnamefont {\ifmmode~\mbox{\c{C}}\else \c{C}\fi{}akmak}},\ }\bibinfo
	{title} {Ergotropy from coherences in an open quantum system},\ \href
	{\doibase 10.1103/PhysRevE.102.042111} {\bibfield  {journal} {\bibinfo
			{journal} {Phys. Rev. E}\ }\textbf {\bibinfo {volume} {102}},\ \bibinfo
		{pages} {042111} (\bibinfo {year} {2020})}\BibitemShut {NoStop}%
	\bibitem [{\citenamefont {Horodecki}\ \emph {et~al.}(2009)\citenamefont
		{Horodecki}, \citenamefont {Horodecki}, \citenamefont {Horodecki},\ and\
		\citenamefont {Horodecki}}]{RevModPhys.81.865}%
	\BibitemOpen
	\bibfield  {author} {\bibinfo {author} {\bibfnamefont {R.}~\bibnamefont
			{Horodecki}}, \bibinfo {author} {\bibfnamefont {P.}~\bibnamefont
			{Horodecki}}, \bibinfo {author} {\bibfnamefont {M.}~\bibnamefont
			{Horodecki}}, \ and\ \bibinfo {author} {\bibfnamefont {K.}~\bibnamefont
			{Horodecki}},\ }\bibinfo {title} {Quantum entanglement},\ \href {\doibase
		10.1103/RevModPhys.81.865} {\bibfield  {journal} {\bibinfo  {journal} {Rev.
				Mod. Phys.}\ }\textbf {\bibinfo {volume} {81}},\ \bibinfo {pages} {865}
		(\bibinfo {year} {2009})}\BibitemShut {NoStop}%
	\bibitem [{\citenamefont {Andolina}\ \emph {et~al.}(2019)\citenamefont
		{Andolina}, \citenamefont {Keck}, \citenamefont {Mari}, \citenamefont
		{Campisi}, \citenamefont {Giovannetti},\ and\ \citenamefont
		{Polini}}]{PhysRevLett.122.047702}%
	\BibitemOpen
	\bibfield  {author} {\bibinfo {author} {\bibfnamefont {G.~M.}\ \bibnamefont
			{Andolina}}, \bibinfo {author} {\bibfnamefont {M.}~\bibnamefont {Keck}},
		\bibinfo {author} {\bibfnamefont {A.}~\bibnamefont {Mari}}, \bibinfo {author}
		{\bibfnamefont {M.}~\bibnamefont {Campisi}}, \bibinfo {author} {\bibfnamefont
			{V.}~\bibnamefont {Giovannetti}}, \ and\ \bibinfo {author} {\bibfnamefont
			{M.}~\bibnamefont {Polini}},\ }\bibinfo {title} {Extractable Work, the Role
		of Correlations, and Asymptotic Freedom in Quantum Batteries},\ \href
	{\doibase 10.1103/PhysRevLett.122.047702} {\bibfield  {journal} {\bibinfo
			{journal} {Phys. Rev. Lett.}\ }\textbf {\bibinfo {volume} {122}},\ \bibinfo
		{pages} {047702} (\bibinfo {year} {2019})}\BibitemShut {NoStop}%
	\bibitem [{\citenamefont {Liu}\ \emph {et~al.}(2021)\citenamefont {Liu},
		\citenamefont {Shi}, \citenamefont {Shi}, \citenamefont {Wang},\ and\
		\citenamefont {Yang}}]{PhysRevB.104.245418}%
	\BibitemOpen
	\bibfield  {author} {\bibinfo {author} {\bibfnamefont {J.-X.}\ \bibnamefont
			{Liu}}, \bibinfo {author} {\bibfnamefont {H.-L.}\ \bibnamefont {Shi}},
		\bibinfo {author} {\bibfnamefont {Y.-H.}\ \bibnamefont {Shi}}, \bibinfo
		{author} {\bibfnamefont {X.-H.}\ \bibnamefont {Wang}}, \ and\ \bibinfo
		{author} {\bibfnamefont {W.-L.}\ \bibnamefont {Yang}},\ }\bibinfo {title}
	{Entanglement and work extraction in the central-spin quantum battery},\
	\href {\doibase 10.1103/PhysRevB.104.245418} {\bibfield  {journal} {\bibinfo
			{journal} {Phys. Rev. B}\ }\textbf {\bibinfo {volume} {104}},\ \bibinfo
		{pages} {245418} (\bibinfo {year} {2021})}\BibitemShut {NoStop}%
	\bibitem [{\citenamefont {Imai}\ \emph {et~al.}(2023)\citenamefont {Imai},
		\citenamefont {G\"uhne},\ and\ \citenamefont
		{Nimmrichter}}]{PhysRevA.107.022215}%
	\BibitemOpen
	\bibfield  {author} {\bibinfo {author} {\bibfnamefont {S.}~\bibnamefont
			{Imai}}, \bibinfo {author} {\bibfnamefont {O.}~\bibnamefont {G\"uhne}}, \
		and\ \bibinfo {author} {\bibfnamefont {S.}~\bibnamefont {Nimmrichter}},\
	}\bibinfo {title} {Work fluctuations and entanglement in quantum batteries},\
	\href {\doibase 10.1103/PhysRevA.107.022215} {\bibfield  {journal} {\bibinfo
			{journal} {Phys. Rev. A}\ }\textbf {\bibinfo {volume} {107}},\ \bibinfo
		{pages} {022215} (\bibinfo {year} {2023})}\BibitemShut {NoStop}%
	\bibitem [{\citenamefont {Ma}\ \emph {et~al.}(2024)\citenamefont {Ma},
		\citenamefont {Xu}, \citenamefont {Li}, \citenamefont {Li},\ and\
		\citenamefont {Zhu}}]{PhysRevA.110.022433}%
	\BibitemOpen
	\bibfield  {author} {\bibinfo {author} {\bibfnamefont {H.-B.}\ \bibnamefont
			{Ma}}, \bibinfo {author} {\bibfnamefont {K.}~\bibnamefont {Xu}}, \bibinfo
		{author} {\bibfnamefont {H.-G.}\ \bibnamefont {Li}}, \bibinfo {author}
		{\bibfnamefont {Z.-G.}\ \bibnamefont {Li}}, \ and\ \bibinfo {author}
		{\bibfnamefont {H.-J.}\ \bibnamefont {Zhu}},\ }\bibinfo {title} {Enhancing
		the charging performance of quantum batteries with the work medium of an
		entangled coupled-cavity array},\ \href {\doibase
		10.1103/PhysRevA.110.022433} {\bibfield  {journal} {\bibinfo  {journal}
			{Phys. Rev. A}\ }\textbf {\bibinfo {volume} {110}},\ \bibinfo {pages}
		{022433} (\bibinfo {year} {2024})}\BibitemShut {NoStop}%
	\bibitem [{\citenamefont {Perarnau-Llobet}\ \emph {et~al.}(2015)\citenamefont
		{Perarnau-Llobet}, \citenamefont {Hovhannisyan}, \citenamefont {Huber},
		\citenamefont {Skrzypczyk}, \citenamefont {Brunner},\ and\ \citenamefont
		{Ac\'{\i}n}}]{PhysRevX.5.041011}%
	\BibitemOpen
	\bibfield  {author} {\bibinfo {author} {\bibfnamefont {M.}~\bibnamefont
			{Perarnau-Llobet}}, \bibinfo {author} {\bibfnamefont {K.~V.}\ \bibnamefont
			{Hovhannisyan}}, \bibinfo {author} {\bibfnamefont {M.}~\bibnamefont {Huber}},
		\bibinfo {author} {\bibfnamefont {P.}~\bibnamefont {Skrzypczyk}}, \bibinfo
		{author} {\bibfnamefont {N.}~\bibnamefont {Brunner}}, \ and\ \bibinfo
		{author} {\bibfnamefont {A.}~\bibnamefont {Ac\'{\i}n}},\ }\bibinfo {title}
	{Extractable Work from Correlations},\ \href {\doibase
		10.1103/PhysRevX.5.041011} {\bibfield  {journal} {\bibinfo  {journal} {Phys.
				Rev. X}\ }\textbf {\bibinfo {volume} {5}},\ \bibinfo {pages} {041011}
		(\bibinfo {year} {2015})}\BibitemShut {NoStop}%
	\bibitem [{\citenamefont {Oppenheim}\ \emph {et~al.}(2002)\citenamefont
		{Oppenheim}, \citenamefont {Horodecki}, \citenamefont {Horodecki},\ and\
		\citenamefont {Horodecki}}]{PhysRevLett.89.180402}%
	\BibitemOpen
	\bibfield  {author} {\bibinfo {author} {\bibfnamefont {J.}~\bibnamefont
			{Oppenheim}}, \bibinfo {author} {\bibfnamefont {M.}~\bibnamefont
			{Horodecki}}, \bibinfo {author} {\bibfnamefont {P.}~\bibnamefont
			{Horodecki}}, \ and\ \bibinfo {author} {\bibfnamefont {R.}~\bibnamefont
			{Horodecki}},\ }\bibinfo {title} {Thermodynamical Approach to Quantifying
		Quantum Correlations},\ \href {\doibase 10.1103/PhysRevLett.89.180402}
	{\bibfield  {journal} {\bibinfo  {journal} {Phys. Rev. Lett.}\ }\textbf
		{\bibinfo {volume} {89}},\ \bibinfo {pages} {180402} (\bibinfo {year}
		{2002})}\BibitemShut {NoStop}%
	\bibitem [{\citenamefont {Allahverdyan}\ \emph {et~al.}(2004)\citenamefont
		{Allahverdyan}, \citenamefont {Balian},\ and\ \citenamefont
		{Nieuwenhuizen}}]{Allahverdyan_2004}%
	\BibitemOpen
	\bibfield  {author} {\bibinfo {author} {\bibfnamefont {A.~E.}\ \bibnamefont
			{Allahverdyan}}, \bibinfo {author} {\bibfnamefont {R.}~\bibnamefont
			{Balian}}, \ and\ \bibinfo {author} {\bibfnamefont {T.~M.}\ \bibnamefont
			{Nieuwenhuizen}},\ }\bibinfo {title} {Maximal work extraction from finite
		quantum systems},\ \href {\doibase 10.1209/epl/i2004-10101-2} {\bibfield
		{journal} {\bibinfo  {journal} {Europhysics Letters}\ }\textbf {\bibinfo
			{volume} {67}},\ \bibinfo {pages} {565} (\bibinfo {year} {2004})}\BibitemShut
	{NoStop}%
	\bibitem [{\citenamefont {Rossini}\ \emph {et~al.}(2020)\citenamefont
		{Rossini}, \citenamefont {Andolina}, \citenamefont {Rosa}, \citenamefont
		{Carrega},\ and\ \citenamefont {Polini}}]{PhysRevLett.125.236402}%
	\BibitemOpen
	\bibfield  {author} {\bibinfo {author} {\bibfnamefont {D.}~\bibnamefont
			{Rossini}}, \bibinfo {author} {\bibfnamefont {G.~M.}\ \bibnamefont
			{Andolina}}, \bibinfo {author} {\bibfnamefont {D.}~\bibnamefont {Rosa}},
		\bibinfo {author} {\bibfnamefont {M.}~\bibnamefont {Carrega}}, \ and\
		\bibinfo {author} {\bibfnamefont {M.}~\bibnamefont {Polini}},\ }\bibinfo
	{title} {Quantum Advantage in the Charging Process of Sachdev-Ye-Kitaev
		Batteries},\ \href {\doibase 10.1103/PhysRevLett.125.236402} {\bibfield
		{journal} {\bibinfo  {journal} {Phys. Rev. Lett.}\ }\textbf {\bibinfo
			{volume} {125}},\ \bibinfo {pages} {236402} (\bibinfo {year}
		{2020})}\BibitemShut {NoStop}%
	\bibitem [{\citenamefont {Carrasco}\ \emph {et~al.}(2022)\citenamefont
		{Carrasco}, \citenamefont {Maze}, \citenamefont {Hermann-Avigliano},\ and\
		\citenamefont {Barra}}]{PhysRevE.105.064119}%
	\BibitemOpen
	\bibfield  {author} {\bibinfo {author} {\bibfnamefont {J.}~\bibnamefont
			{Carrasco}}, \bibinfo {author} {\bibfnamefont {J.~R.}\ \bibnamefont {Maze}},
		\bibinfo {author} {\bibfnamefont {C.}~\bibnamefont {Hermann-Avigliano}}, \
		and\ \bibinfo {author} {\bibfnamefont {F.}~\bibnamefont {Barra}},\ }\bibinfo
	{title} {Collective enhancement in dissipative quantum batteries},\ \href
	{\doibase 10.1103/PhysRevE.105.064119} {\bibfield  {journal} {\bibinfo
			{journal} {Phys. Rev. E}\ }\textbf {\bibinfo {volume} {105}},\ \bibinfo
		{pages} {064119} (\bibinfo {year} {2022})}\BibitemShut {NoStop}%
	\bibitem [{\citenamefont {Elghaayda}\ \emph {et~al.}(2025)\citenamefont
		{Elghaayda}, \citenamefont {Ali}, \citenamefont {Al-Kuwari}, \citenamefont
		{Czerwinski}, \citenamefont {Mansour},\ and\ \citenamefont
		{Haddadi}}]{10.1002/qute.202400651}%
	\BibitemOpen
	\bibfield  {author} {\bibinfo {author} {\bibfnamefont {S.}~\bibnamefont
			{Elghaayda}}, \bibinfo {author} {\bibfnamefont {A.}~\bibnamefont {Ali}},
		\bibinfo {author} {\bibfnamefont {S.}~\bibnamefont {Al-Kuwari}}, \bibinfo
		{author} {\bibfnamefont {A.}~\bibnamefont {Czerwinski}}, \bibinfo {author}
		{\bibfnamefont {M.}~\bibnamefont {Mansour}}, \ and\ \bibinfo {author}
		{\bibfnamefont {S.}~\bibnamefont {Haddadi}},\ }\bibinfo {title} {Performance
		of a Superconducting Quantum Battery},\ \href {\doibase
		https://doi.org/10.1002/qute.202400651} {\bibfield  {journal} {\bibinfo
			{journal} {Advanced Quantum Technologies}\ }\textbf {\bibinfo {volume} {8}},\
		\bibinfo {pages} {2400651} (\bibinfo {year} {2025})}\BibitemShut {NoStop}%
	\bibitem [{\citenamefont {Ferraro}\ \emph {et~al.}(2018)\citenamefont
		{Ferraro}, \citenamefont {Campisi}, \citenamefont {Andolina}, \citenamefont
		{Pellegrini},\ and\ \citenamefont {Polini}}]{PhysRevLett.120.117702}%
	\BibitemOpen
	\bibfield  {author} {\bibinfo {author} {\bibfnamefont {D.}~\bibnamefont
			{Ferraro}}, \bibinfo {author} {\bibfnamefont {M.}~\bibnamefont {Campisi}},
		\bibinfo {author} {\bibfnamefont {G.~M.}\ \bibnamefont {Andolina}}, \bibinfo
		{author} {\bibfnamefont {V.}~\bibnamefont {Pellegrini}}, \ and\ \bibinfo
		{author} {\bibfnamefont {M.}~\bibnamefont {Polini}},\ }\bibinfo {title}
	{High-Power Collective Charging of a Solid-State Quantum Battery},\ \href
	{\doibase 10.1103/PhysRevLett.120.117702} {\bibfield  {journal} {\bibinfo
			{journal} {Phys. Rev. Lett.}\ }\textbf {\bibinfo {volume} {120}},\ \bibinfo
		{pages} {117702} (\bibinfo {year} {2018})}\BibitemShut {NoStop}%
	\bibitem [{\citenamefont {Dou}\ \emph {et~al.}(2022{\natexlab{a}})\citenamefont
		{Dou}, \citenamefont {Lu}, \citenamefont {Wang},\ and\ \citenamefont
		{Sun}}]{PhysRevB.105.115405}%
	\BibitemOpen
	\bibfield  {author} {\bibinfo {author} {\bibfnamefont {F.-Q.}\ \bibnamefont
			{Dou}}, \bibinfo {author} {\bibfnamefont {Y.-Q.}\ \bibnamefont {Lu}},
		\bibinfo {author} {\bibfnamefont {Y.-J.}\ \bibnamefont {Wang}}, \ and\
		\bibinfo {author} {\bibfnamefont {J.-A.}\ \bibnamefont {Sun}},\ }\bibinfo
	{title} {Extended Dicke quantum battery with interatomic interactions and
		driving field},\ \href {\doibase 10.1103/PhysRevB.105.115405} {\bibfield
		{journal} {\bibinfo  {journal} {Phys. Rev. B}\ }\textbf {\bibinfo {volume}
			{105}},\ \bibinfo {pages} {115405} (\bibinfo {year}
		{2022}{\natexlab{a}})}\BibitemShut {NoStop}%
	\bibitem [{\citenamefont {Joshi}\ and\ \citenamefont
		{Mahesh}(2022)}]{PhysRevA.106.042601}%
	\BibitemOpen
	\bibfield  {author} {\bibinfo {author} {\bibfnamefont {J.}~\bibnamefont
			{Joshi}}\ and\ \bibinfo {author} {\bibfnamefont {T.~S.}\ \bibnamefont
			{Mahesh}},\ }\bibinfo {title} {Experimental investigation of a quantum
		battery using star-topology NMR spin systems},\ \href {\doibase
		10.1103/PhysRevA.106.042601} {\bibfield  {journal} {\bibinfo  {journal}
			{Phys. Rev. A}\ }\textbf {\bibinfo {volume} {106}},\ \bibinfo {pages}
		{042601} (\bibinfo {year} {2022})}\BibitemShut {NoStop}%
	\bibitem [{\citenamefont {Pokhrel}\ and\ \citenamefont
		{Gea-Banacloche}(2025)}]{PhysRevLett.134.130401}%
	\BibitemOpen
	\bibfield  {author} {\bibinfo {author} {\bibfnamefont {S.}~\bibnamefont
			{Pokhrel}}\ and\ \bibinfo {author} {\bibfnamefont {J.}~\bibnamefont
			{Gea-Banacloche}},\ }\bibinfo {title} {Large Collective Power Enhancement in
		Dissipative Charging of a Quantum Battery},\ \href {\doibase
		10.1103/PhysRevLett.134.130401} {\bibfield  {journal} {\bibinfo  {journal}
			{Phys. Rev. Lett.}\ }\textbf {\bibinfo {volume} {134}},\ \bibinfo {pages}
		{130401} (\bibinfo {year} {2025})}\BibitemShut {NoStop}%
	\bibitem [{\citenamefont {Santos}\ \emph {et~al.}(2019)\citenamefont {Santos},
		\citenamefont {\ifmmode~\mbox{\c{C}}\else \c{C}\fi{}akmak}, \citenamefont
		{Campbell},\ and\ \citenamefont {Zinner}}]{PhysRevE.100.032107}%
	\BibitemOpen
	\bibfield  {author} {\bibinfo {author} {\bibfnamefont {A.~C.}\ \bibnamefont
			{Santos}}, \bibinfo {author} {\bibfnamefont {B.~i. e. i. f. m.~c.}\
			\bibnamefont {\ifmmode~\mbox{\c{C}}\else \c{C}\fi{}akmak}}, \bibinfo {author}
		{\bibfnamefont {S.}~\bibnamefont {Campbell}}, \ and\ \bibinfo {author}
		{\bibfnamefont {N.~T.}\ \bibnamefont {Zinner}},\ }\bibinfo {title} {Stable
		adiabatic quantum batteries},\ \href {\doibase 10.1103/PhysRevE.100.032107}
	{\bibfield  {journal} {\bibinfo  {journal} {Phys. Rev. E}\ }\textbf {\bibinfo
			{volume} {100}},\ \bibinfo {pages} {032107} (\bibinfo {year}
		{2019})}\BibitemShut {NoStop}%
	\bibitem [{\citenamefont {Fasihi}\ \emph {et~al.}(2025)\citenamefont {Fasihi},
		\citenamefont {Jafarzadeh~Bahrbeig}, \citenamefont {Mojaveri},\ and\
		\citenamefont {Haji~Mohammadzadeh}}]{6c73-ll23}%
	\BibitemOpen
	\bibfield  {author} {\bibinfo {author} {\bibfnamefont {M.~A.}\ \bibnamefont
			{Fasihi}}, \bibinfo {author} {\bibfnamefont {R.}~\bibnamefont
			{Jafarzadeh~Bahrbeig}}, \bibinfo {author} {\bibfnamefont {B.}~\bibnamefont
			{Mojaveri}}, \ and\ \bibinfo {author} {\bibfnamefont {R.}~\bibnamefont
			{Haji~Mohammadzadeh}},\ }\bibinfo {title} {Fast stable adiabatic charging of
		open quantum batteries},\ \href {\doibase 10.1103/6c73-ll23} {\bibfield
		{journal} {\bibinfo  {journal} {Phys. Rev. E}\ }\textbf {\bibinfo {volume}
			{112}},\ \bibinfo {pages} {024117} (\bibinfo {year} {2025})}\BibitemShut
	{NoStop}%
	\bibitem [{\citenamefont {Dou}\ \emph {et~al.}(2022{\natexlab{b}})\citenamefont
		{Dou}, \citenamefont {Wang},\ and\ \citenamefont {Sun}}]{dou2022highly}%
	\BibitemOpen
	\bibfield  {author} {\bibinfo {author} {\bibfnamefont {F.-Q.}\ \bibnamefont
			{Dou}}, \bibinfo {author} {\bibfnamefont {Y.-J.}\ \bibnamefont {Wang}}, \
		and\ \bibinfo {author} {\bibfnamefont {J.-A.}\ \bibnamefont {Sun}},\
	}\bibinfo {title} {Highly efficient charging and discharging of three-level
		quantum batteries through shortcuts to adiabaticity},\ \href@noop {}
	{\bibfield  {journal} {\bibinfo  {journal} {Frontiers of Physics}\ }\textbf
		{\bibinfo {volume} {17}},\ \bibinfo {pages} {31503} (\bibinfo {year}
		{2022}{\natexlab{b}})}\BibitemShut {NoStop}%
	\bibitem [{\citenamefont {Yang}\ \emph
		{et~al.}(2024{\natexlab{b}})\citenamefont {Yang}, \citenamefont {Yang},\ and\
		\citenamefont {Dou}}]{PhysRevB.109.235432}%
	\BibitemOpen
	\bibfield  {author} {\bibinfo {author} {\bibfnamefont {D.-L.}\ \bibnamefont
			{Yang}}, \bibinfo {author} {\bibfnamefont {F.-M.}\ \bibnamefont {Yang}}, \
		and\ \bibinfo {author} {\bibfnamefont {F.-Q.}\ \bibnamefont {Dou}},\
	}\bibinfo {title} {Three-level Dicke quantum battery},\ \href {\doibase
		10.1103/PhysRevB.109.235432} {\bibfield  {journal} {\bibinfo  {journal}
			{Phys. Rev. B}\ }\textbf {\bibinfo {volume} {109}},\ \bibinfo {pages}
		{235432} (\bibinfo {year} {2024}{\natexlab{b}})}\BibitemShut {NoStop}%
	\bibitem [{\citenamefont {Erdman}\ \emph {et~al.}(2024)\citenamefont {Erdman},
		\citenamefont {Andolina}, \citenamefont {Giovannetti},\ and\ \citenamefont
		{No\'e}}]{PhysRevLett.133.243602}%
	\BibitemOpen
	\bibfield  {author} {\bibinfo {author} {\bibfnamefont {P.~A.}\ \bibnamefont
			{Erdman}}, \bibinfo {author} {\bibfnamefont {G.~M.}\ \bibnamefont
			{Andolina}}, \bibinfo {author} {\bibfnamefont {V.}~\bibnamefont
			{Giovannetti}}, \ and\ \bibinfo {author} {\bibfnamefont {F.}~\bibnamefont
			{No\'e}},\ }\bibinfo {title} {Reinforcement Learning Optimization of the
		Charging of a Dicke Quantum Battery},\ \href {\doibase
		10.1103/PhysRevLett.133.243602} {\bibfield  {journal} {\bibinfo  {journal}
			{Phys. Rev. Lett.}\ }\textbf {\bibinfo {volume} {133}},\ \bibinfo {pages}
		{243602} (\bibinfo {year} {2024})}\BibitemShut {NoStop}%
	\bibitem [{\citenamefont {Seidov}\ and\ \citenamefont
		{Mukhin}(2024)}]{PhysRevA.109.022210}%
	\BibitemOpen
	\bibfield  {author} {\bibinfo {author} {\bibfnamefont {S.~S.}\ \bibnamefont
			{Seidov}}\ and\ \bibinfo {author} {\bibfnamefont {S.~I.}\ \bibnamefont
			{Mukhin}},\ }\bibinfo {title} {Quantum Dicke battery supercharging in the
		bound-luminosity state},\ \href {\doibase 10.1103/PhysRevA.109.022210}
	{\bibfield  {journal} {\bibinfo  {journal} {Phys. Rev. A}\ }\textbf {\bibinfo
			{volume} {109}},\ \bibinfo {pages} {022210} (\bibinfo {year}
		{2024})}\BibitemShut {NoStop}%
	\bibitem [{\citenamefont {Gao}\ \emph {et~al.}(2022)\citenamefont {Gao},
		\citenamefont {Cheng}, \citenamefont {He}, \citenamefont {Mondaini},
		\citenamefont {Guan},\ and\ \citenamefont {Lin}}]{PhysRevResearch.4.043150}%
	\BibitemOpen
	\bibfield  {author} {\bibinfo {author} {\bibfnamefont {L.}~\bibnamefont
			{Gao}}, \bibinfo {author} {\bibfnamefont {C.}~\bibnamefont {Cheng}}, \bibinfo
		{author} {\bibfnamefont {W.-B.}\ \bibnamefont {He}}, \bibinfo {author}
		{\bibfnamefont {R.}~\bibnamefont {Mondaini}}, \bibinfo {author}
		{\bibfnamefont {X.-W.}\ \bibnamefont {Guan}}, \ and\ \bibinfo {author}
		{\bibfnamefont {H.-Q.}\ \bibnamefont {Lin}},\ }\bibinfo {title} {Scaling of
		energy and power in a large quantum battery-charger model},\ \href {\doibase
		10.1103/PhysRevResearch.4.043150} {\bibfield  {journal} {\bibinfo  {journal}
			{Phys. Rev. Res.}\ }\textbf {\bibinfo {volume} {4}},\ \bibinfo {pages}
		{043150} (\bibinfo {year} {2022})}\BibitemShut {NoStop}%
	\bibitem [{\citenamefont {Sun}\ and\ \citenamefont
		{Ye}(2020)}]{PhysRevLett.124.244101}%
	\BibitemOpen
	\bibfield  {author} {\bibinfo {author} {\bibfnamefont {F.}~\bibnamefont
			{Sun}}\ and\ \bibinfo {author} {\bibfnamefont {J.}~\bibnamefont {Ye}},\
	}\bibinfo {title} {Periodic Table of the Ordinary and Supersymmetric
		Sachdev-Ye-Kitaev Models},\ \href {\doibase 10.1103/PhysRevLett.124.244101}
	{\bibfield  {journal} {\bibinfo  {journal} {Phys. Rev. Lett.}\ }\textbf
		{\bibinfo {volume} {124}},\ \bibinfo {pages} {244101} (\bibinfo {year}
		{2020})}\BibitemShut {NoStop}%
	\bibitem [{\citenamefont {Le}\ \emph {et~al.}(2018)\citenamefont {Le},
		\citenamefont {Levinsen}, \citenamefont {Modi}, \citenamefont {Parish},\ and\
		\citenamefont {Pollock}}]{PhysRevA.97.022106}%
	\BibitemOpen
	\bibfield  {author} {\bibinfo {author} {\bibfnamefont {T.~P.}\ \bibnamefont
			{Le}}, \bibinfo {author} {\bibfnamefont {J.}~\bibnamefont {Levinsen}},
		\bibinfo {author} {\bibfnamefont {K.}~\bibnamefont {Modi}}, \bibinfo {author}
		{\bibfnamefont {M.~M.}\ \bibnamefont {Parish}}, \ and\ \bibinfo {author}
		{\bibfnamefont {F.~A.}\ \bibnamefont {Pollock}},\ }\bibinfo {title}
	{Spin-chain model of a many-body quantum battery},\ \href {\doibase
		10.1103/PhysRevA.97.022106} {\bibfield  {journal} {\bibinfo  {journal} {Phys.
				Rev. A}\ }\textbf {\bibinfo {volume} {97}},\ \bibinfo {pages} {022106}
		(\bibinfo {year} {2018})}\BibitemShut {NoStop}%
	\bibitem [{\citenamefont {Grazi}\ \emph {et~al.}(2024)\citenamefont {Grazi},
		\citenamefont {Sacco~Shaikh}, \citenamefont {Sassetti}, \citenamefont
		{Traverso~Ziani},\ and\ \citenamefont {Ferraro}}]{PhysRevLett.133.197001}%
	\BibitemOpen
	\bibfield  {author} {\bibinfo {author} {\bibfnamefont {R.}~\bibnamefont
			{Grazi}}, \bibinfo {author} {\bibfnamefont {D.}~\bibnamefont {Sacco~Shaikh}},
		\bibinfo {author} {\bibfnamefont {M.}~\bibnamefont {Sassetti}}, \bibinfo
		{author} {\bibfnamefont {N.}~\bibnamefont {Traverso~Ziani}}, \ and\ \bibinfo
		{author} {\bibfnamefont {D.}~\bibnamefont {Ferraro}},\ }\bibinfo {title}
	{Controlling Energy Storage Crossing Quantum Phase Transitions in an
		Integrable Spin Quantum Battery},\ \href {\doibase
		10.1103/PhysRevLett.133.197001} {\bibfield  {journal} {\bibinfo  {journal}
			{Phys. Rev. Lett.}\ }\textbf {\bibinfo {volume} {133}},\ \bibinfo {pages}
		{197001} (\bibinfo {year} {2024})}\BibitemShut {NoStop}%
	\bibitem [{\citenamefont {Dou}\ \emph {et~al.}(2022{\natexlab{c}})\citenamefont
		{Dou}, \citenamefont {Zhou},\ and\ \citenamefont
		{Sun}}]{PhysRevA.106.032212}%
	\BibitemOpen
	\bibfield  {author} {\bibinfo {author} {\bibfnamefont {F.-Q.}\ \bibnamefont
			{Dou}}, \bibinfo {author} {\bibfnamefont {H.}~\bibnamefont {Zhou}}, \ and\
		\bibinfo {author} {\bibfnamefont {J.-A.}\ \bibnamefont {Sun}},\ }\bibinfo
	{title} {Cavity Heisenberg-spin-chain quantum battery},\ \href {\doibase
		10.1103/PhysRevA.106.032212} {\bibfield  {journal} {\bibinfo  {journal}
			{Phys. Rev. A}\ }\textbf {\bibinfo {volume} {106}},\ \bibinfo {pages}
		{032212} (\bibinfo {year} {2022}{\natexlab{c}})}\BibitemShut {NoStop}%
	\bibitem [{\citenamefont {Zhao}\ \emph {et~al.}(2022)\citenamefont {Zhao},
		\citenamefont {Dou},\ and\ \citenamefont {Zhao}}]{PhysRevResearch.4.013172}%
	\BibitemOpen
	\bibfield  {author} {\bibinfo {author} {\bibfnamefont {F.}~\bibnamefont
			{Zhao}}, \bibinfo {author} {\bibfnamefont {F.-Q.}\ \bibnamefont {Dou}}, \
		and\ \bibinfo {author} {\bibfnamefont {Q.}~\bibnamefont {Zhao}},\ }\bibinfo
	{title} {Charging performance of the Su-Schrieffer-Heeger quantum battery},\
	\href {\doibase 10.1103/PhysRevResearch.4.013172} {\bibfield  {journal}
		{\bibinfo  {journal} {Phys. Rev. Res.}\ }\textbf {\bibinfo {volume} {4}},\
		\bibinfo {pages} {013172} (\bibinfo {year} {2022})}\BibitemShut {NoStop}%
	\bibitem [{\citenamefont {Rodr\'{\i}guez}\ \emph {et~al.}(2023)\citenamefont
		{Rodr\'{\i}guez}, \citenamefont {Ahmadi}, \citenamefont {Mazurek},
		\citenamefont {Barzanjeh}, \citenamefont {Alicki},\ and\ \citenamefont
		{Horodecki}}]{PhysRevA.107.042419}%
	\BibitemOpen
	\bibfield  {author} {\bibinfo {author} {\bibfnamefont {R.~R.}\ \bibnamefont
			{Rodr\'{\i}guez}}, \bibinfo {author} {\bibfnamefont {B.}~\bibnamefont
			{Ahmadi}}, \bibinfo {author} {\bibfnamefont {P.}~\bibnamefont {Mazurek}},
		\bibinfo {author} {\bibfnamefont {S.}~\bibnamefont {Barzanjeh}}, \bibinfo
		{author} {\bibfnamefont {R.}~\bibnamefont {Alicki}}, \ and\ \bibinfo {author}
		{\bibfnamefont {P.}~\bibnamefont {Horodecki}},\ }\bibinfo {title} {Catalysis
		in charging quantum batteries},\ \href {\doibase 10.1103/PhysRevA.107.042419}
	{\bibfield  {journal} {\bibinfo  {journal} {Phys. Rev. A}\ }\textbf {\bibinfo
			{volume} {107}},\ \bibinfo {pages} {042419} (\bibinfo {year}
		{2023})}\BibitemShut {NoStop}%
	\bibitem [{\citenamefont {Andolina}\ \emph {et~al.}(2018)\citenamefont
		{Andolina}, \citenamefont {Farina}, \citenamefont {Mari}, \citenamefont
		{Pellegrini}, \citenamefont {Giovannetti},\ and\ \citenamefont
		{Polini}}]{PhysRevB.98.205423}%
	\BibitemOpen
	\bibfield  {author} {\bibinfo {author} {\bibfnamefont {G.~M.}\ \bibnamefont
			{Andolina}}, \bibinfo {author} {\bibfnamefont {D.}~\bibnamefont {Farina}},
		\bibinfo {author} {\bibfnamefont {A.}~\bibnamefont {Mari}}, \bibinfo {author}
		{\bibfnamefont {V.}~\bibnamefont {Pellegrini}}, \bibinfo {author}
		{\bibfnamefont {V.}~\bibnamefont {Giovannetti}}, \ and\ \bibinfo {author}
		{\bibfnamefont {M.}~\bibnamefont {Polini}},\ }\bibinfo {title}
	{Charger-mediated energy transfer in exactly solvable models for quantum
		batteries},\ \href {\doibase 10.1103/PhysRevB.98.205423} {\bibfield
		{journal} {\bibinfo  {journal} {Phys. Rev. B}\ }\textbf {\bibinfo {volume}
			{98}},\ \bibinfo {pages} {205423} (\bibinfo {year} {2018})}\BibitemShut
	{NoStop}%
	\bibitem [{\citenamefont {Farina}\ \emph {et~al.}(2019)\citenamefont {Farina},
		\citenamefont {Andolina}, \citenamefont {Mari}, \citenamefont {Polini},\ and\
		\citenamefont {Giovannetti}}]{PhysRevB.99.035421}%
	\BibitemOpen
	\bibfield  {author} {\bibinfo {author} {\bibfnamefont {D.}~\bibnamefont
			{Farina}}, \bibinfo {author} {\bibfnamefont {G.~M.}\ \bibnamefont
			{Andolina}}, \bibinfo {author} {\bibfnamefont {A.}~\bibnamefont {Mari}},
		\bibinfo {author} {\bibfnamefont {M.}~\bibnamefont {Polini}}, \ and\ \bibinfo
		{author} {\bibfnamefont {V.}~\bibnamefont {Giovannetti}},\ }\bibinfo {title}
	{Charger-mediated energy transfer for quantum batteries: An open-system
		approach},\ \href {\doibase 10.1103/PhysRevB.99.035421} {\bibfield  {journal}
		{\bibinfo  {journal} {Phys. Rev. B}\ }\textbf {\bibinfo {volume} {99}},\
		\bibinfo {pages} {035421} (\bibinfo {year} {2019})}\BibitemShut {NoStop}%
	\bibitem [{\citenamefont {Grebenkov}\ and\ \citenamefont
		{Skvortsov}(2020)}]{Grebenkov_2020}%
	\BibitemOpen
	\bibfield  {author} {\bibinfo {author} {\bibfnamefont {D.~S.}\ \bibnamefont
			{Grebenkov}}\ and\ \bibinfo {author} {\bibfnamefont {A.~T.}\ \bibnamefont
			{Skvortsov}},\ }\bibinfo {title} {Mean first-passage time to a small
		absorbing target in an elongated planar domain},\ \href {\doibase
		10.1088/1367-2630/abc91f} {\bibfield  {journal} {\bibinfo  {journal} {New
				Journal of Physics}\ }\textbf {\bibinfo {volume} {22}},\ \bibinfo {pages}
		{113024} (\bibinfo {year} {2020})}\BibitemShut {NoStop}%
	\bibitem [{\citenamefont {Wang}\ \emph {et~al.}(2024)\citenamefont {Wang},
		\citenamefont {Liu}, \citenamefont {Wu}, \citenamefont {Fan},\ and\
		\citenamefont {Liu}}]{PhysRevA.110.062204}%
	\BibitemOpen
	\bibfield  {author} {\bibinfo {author} {\bibfnamefont {L.}~\bibnamefont
			{Wang}}, \bibinfo {author} {\bibfnamefont {S.-Q.}\ \bibnamefont {Liu}},
		\bibinfo {author} {\bibfnamefont {F.-l.}\ \bibnamefont {Wu}}, \bibinfo
		{author} {\bibfnamefont {H.}~\bibnamefont {Fan}}, \ and\ \bibinfo {author}
		{\bibfnamefont {S.-Y.}\ \bibnamefont {Liu}},\ }\bibinfo {title}
	{Cavity-optomechanical quantum battery},\ \href {\doibase
		10.1103/PhysRevA.110.062204} {\bibfield  {journal} {\bibinfo  {journal}
			{Phys. Rev. A}\ }\textbf {\bibinfo {volume} {110}},\ \bibinfo {pages}
		{062204} (\bibinfo {year} {2024})}\BibitemShut {NoStop}%
	\bibitem [{\citenamefont {Qi}\ and\ \citenamefont
		{Jing}(2021)}]{PhysRevA.104.032606}%
	\BibitemOpen
	\bibfield  {author} {\bibinfo {author} {\bibfnamefont {S.-f.}\ \bibnamefont
			{Qi}}\ and\ \bibinfo {author} {\bibfnamefont {J.}~\bibnamefont {Jing}},\
	}\bibinfo {title} {Magnon-mediated quantum battery under systematic errors},\
	\href {\doibase 10.1103/PhysRevA.104.032606} {\bibfield  {journal} {\bibinfo
			{journal} {Phys. Rev. A}\ }\textbf {\bibinfo {volume} {104}},\ \bibinfo
		{pages} {032606} (\bibinfo {year} {2021})}\BibitemShut {NoStop}%
	\bibitem [{\citenamefont {Chiribella}\ \emph {et~al.}(2021)\citenamefont
		{Chiribella}, \citenamefont {Yang},\ and\ \citenamefont
		{Renner}}]{PhysRevX.11.021014}%
	\BibitemOpen
	\bibfield  {author} {\bibinfo {author} {\bibfnamefont {G.}~\bibnamefont
			{Chiribella}}, \bibinfo {author} {\bibfnamefont {Y.}~\bibnamefont {Yang}}, \
		and\ \bibinfo {author} {\bibfnamefont {R.}~\bibnamefont {Renner}},\ }\bibinfo
	{title} {Fundamental Energy Requirement of Reversible Quantum Operations},\
	\href {\doibase 10.1103/PhysRevX.11.021014} {\bibfield  {journal} {\bibinfo
			{journal} {Phys. Rev. X}\ }\textbf {\bibinfo {volume} {11}},\ \bibinfo
		{pages} {021014} (\bibinfo {year} {2021})}\BibitemShut {NoStop}%
	\bibitem [{\citenamefont {Niedenzu}\ \emph {et~al.}(2018)\citenamefont
		{Niedenzu}, \citenamefont {Mukherjee}, \citenamefont {Ghosh}, \citenamefont
		{Kofman},\ and\ \citenamefont {Kurizki}}]{niedenzu2018quantum}%
	\BibitemOpen
	\bibfield  {author} {\bibinfo {author} {\bibfnamefont {W.}~\bibnamefont
			{Niedenzu}}, \bibinfo {author} {\bibfnamefont {V.}~\bibnamefont {Mukherjee}},
		\bibinfo {author} {\bibfnamefont {A.}~\bibnamefont {Ghosh}}, \bibinfo
		{author} {\bibfnamefont {A.~G.}\ \bibnamefont {Kofman}}, \ and\ \bibinfo
		{author} {\bibfnamefont {G.}~\bibnamefont {Kurizki}},\ }\bibinfo {title}
	{Quantum engine efficiency bound beyond the second law of thermodynamics},\
	\href@noop {} {\bibfield  {journal} {\bibinfo  {journal} {Nature
				communications}\ }\textbf {\bibinfo {volume} {9}},\ \bibinfo {pages} {165}
		(\bibinfo {year} {2018})}\BibitemShut {NoStop}%
	\bibitem [{\citenamefont {Bhattacharjee}\ and\ \citenamefont
		{Dutta}(2021)}]{bhattacharjee2021quantum}%
	\BibitemOpen
	\bibfield  {author} {\bibinfo {author} {\bibfnamefont {S.}~\bibnamefont
			{Bhattacharjee}}\ and\ \bibinfo {author} {\bibfnamefont {A.}~\bibnamefont
			{Dutta}},\ }\bibinfo {title} {Quantum thermal machines and batteries},\
	\href@noop {} {\bibfield  {journal} {\bibinfo  {journal} {The European
				Physical Journal B}\ }\textbf {\bibinfo {volume} {94}},\ \bibinfo {pages}
		{239} (\bibinfo {year} {2021})}\BibitemShut {NoStop}%
	\bibitem [{\citenamefont {Giorgi}\ and\ \citenamefont
		{Campbell}(2015)}]{Giorgi_2015}%
	\BibitemOpen
	\bibfield  {author} {\bibinfo {author} {\bibfnamefont {G.~L.}\ \bibnamefont
			{Giorgi}}\ and\ \bibinfo {author} {\bibfnamefont {S.}~\bibnamefont
			{Campbell}},\ }\bibinfo {title} {Correlation approach to work extraction from
		finite quantum systems},\ \href {\doibase 10.1088/0953-4075/48/3/035501}
	{\bibfield  {journal} {\bibinfo  {journal} {Journal of Physics B: Atomic,
				Molecular and Optical Physics}\ }\textbf {\bibinfo {volume} {48}},\ \bibinfo
		{pages} {035501} (\bibinfo {year} {2015})}\BibitemShut {NoStop}%
	\bibitem [{\citenamefont {Caravelli}\ \emph {et~al.}(2020)\citenamefont
		{Caravelli}, \citenamefont {Coulter-De~Wit}, \citenamefont
		{Garc\'{\i}a-Pintos},\ and\ \citenamefont
		{Hamma}}]{PhysRevResearch.2.023095}%
	\BibitemOpen
	\bibfield  {author} {\bibinfo {author} {\bibfnamefont {F.}~\bibnamefont
			{Caravelli}}, \bibinfo {author} {\bibfnamefont {G.}~\bibnamefont
			{Coulter-De~Wit}}, \bibinfo {author} {\bibfnamefont {L.~P.}\ \bibnamefont
			{Garc\'{\i}a-Pintos}}, \ and\ \bibinfo {author} {\bibfnamefont
			{A.}~\bibnamefont {Hamma}},\ }\bibinfo {title} {Random quantum batteries},\
	\href {\doibase 10.1103/PhysRevResearch.2.023095} {\bibfield  {journal}
		{\bibinfo  {journal} {Phys. Rev. Res.}\ }\textbf {\bibinfo {volume} {2}},\
		\bibinfo {pages} {023095} (\bibinfo {year} {2020})}\BibitemShut {NoStop}%
	\bibitem [{\citenamefont {Stevens}\ \emph {et~al.}(2022)\citenamefont
		{Stevens}, \citenamefont {Szombati}, \citenamefont {Maffei}, \citenamefont
		{Elouard}, \citenamefont {Assouly}, \citenamefont {Cottet}, \citenamefont
		{Dassonneville}, \citenamefont {Ficheux}, \citenamefont {Zeppetzauer},
		\citenamefont {Bienfait}, \citenamefont {Jordan}, \citenamefont
		{Auff\`eves},\ and\ \citenamefont {Huard}}]{PhysRevLett.129.110601}%
	\BibitemOpen
	\bibfield  {author} {\bibinfo {author} {\bibfnamefont {J.}~\bibnamefont
			{Stevens}}, \bibinfo {author} {\bibfnamefont {D.}~\bibnamefont {Szombati}},
		\bibinfo {author} {\bibfnamefont {M.}~\bibnamefont {Maffei}}, \bibinfo
		{author} {\bibfnamefont {C.}~\bibnamefont {Elouard}}, \bibinfo {author}
		{\bibfnamefont {R.}~\bibnamefont {Assouly}}, \bibinfo {author} {\bibfnamefont
			{N.}~\bibnamefont {Cottet}}, \bibinfo {author} {\bibfnamefont
			{R.}~\bibnamefont {Dassonneville}}, \bibinfo {author} {\bibfnamefont
			{Q.}~\bibnamefont {Ficheux}}, \bibinfo {author} {\bibfnamefont
			{S.}~\bibnamefont {Zeppetzauer}}, \bibinfo {author} {\bibfnamefont
			{A.}~\bibnamefont {Bienfait}}, \bibinfo {author} {\bibfnamefont {A.~N.}\
			\bibnamefont {Jordan}}, \bibinfo {author} {\bibfnamefont {A.}~\bibnamefont
			{Auff\`eves}}, \ and\ \bibinfo {author} {\bibfnamefont {B.}~\bibnamefont
			{Huard}},\ }\bibinfo {title} {Energetics of a Single Qubit Gate},\ \href
	{\doibase 10.1103/PhysRevLett.129.110601} {\bibfield  {journal} {\bibinfo
			{journal} {Phys. Rev. Lett.}\ }\textbf {\bibinfo {volume} {129}},\ \bibinfo
		{pages} {110601} (\bibinfo {year} {2022})}\BibitemShut {NoStop}%
	\bibitem [{\citenamefont {Parrondo}\ \emph {et~al.}(2015)\citenamefont
		{Parrondo}, \citenamefont {Horowitz},\ and\ \citenamefont
		{Sagawa}}]{parrondo2015thermodynamics}%
	\BibitemOpen
	\bibfield  {author} {\bibinfo {author} {\bibfnamefont {J.~M.}\ \bibnamefont
			{Parrondo}}, \bibinfo {author} {\bibfnamefont {J.~M.}\ \bibnamefont
			{Horowitz}}, \ and\ \bibinfo {author} {\bibfnamefont {T.}~\bibnamefont
			{Sagawa}},\ }\bibinfo {title} {Thermodynamics of information},\ \href@noop {}
	{\bibfield  {journal} {\bibinfo  {journal} {Nature physics}\ }\textbf
		{\bibinfo {volume} {11}},\ \bibinfo {pages} {131} (\bibinfo {year}
		{2015})}\BibitemShut {NoStop}%
	\bibitem [{\citenamefont {Millen}\ and\ \citenamefont
		{Xuereb}(2016)}]{Millen_2016}%
	\BibitemOpen
	\bibfield  {author} {\bibinfo {author} {\bibfnamefont {J.}~\bibnamefont
			{Millen}}\ and\ \bibinfo {author} {\bibfnamefont {A.}~\bibnamefont
			{Xuereb}},\ }\bibinfo {title} {Perspective on quantum thermodynamics},\ \href
	{\doibase 10.1088/1367-2630/18/1/011002} {\bibfield  {journal} {\bibinfo
			{journal} {New Journal of Physics}\ }\textbf {\bibinfo {volume} {18}},\
		\bibinfo {pages} {011002} (\bibinfo {year} {2016})}\BibitemShut {NoStop}%
	\bibitem [{\citenamefont {Auff\`eves}(2022)}]{PRXQuantum.3.020101}%
	\BibitemOpen
	\bibfield  {author} {\bibinfo {author} {\bibfnamefont {A.}~\bibnamefont
			{Auff\`eves}},\ }\bibinfo {title} {Quantum Technologies Need a Quantum Energy
		Initiative},\ \href {\doibase 10.1103/PRXQuantum.3.020101} {\bibfield
		{journal} {\bibinfo  {journal} {PRX Quantum}\ }\textbf {\bibinfo {volume}
			{3}},\ \bibinfo {pages} {020101} (\bibinfo {year} {2022})}\BibitemShut
	{NoStop}%
	\bibitem [{\citenamefont {Horodecki}\ and\ \citenamefont
		{Oppenheim}(2013)}]{horodecki2013fundamental}%
	\BibitemOpen
	\bibfield  {author} {\bibinfo {author} {\bibfnamefont {M.}~\bibnamefont
			{Horodecki}}\ and\ \bibinfo {author} {\bibfnamefont {J.}~\bibnamefont
			{Oppenheim}},\ }\bibinfo {title} {Fundamental limitations for quantum and
		nanoscale thermodynamics},\ \href@noop {} {\bibfield  {journal} {\bibinfo
			{journal} {Nature communications}\ }\textbf {\bibinfo {volume} {4}},\
		\bibinfo {pages} {2059} (\bibinfo {year} {2013})}\BibitemShut {NoStop}%
	\bibitem [{\citenamefont {Skrzypczyk}\ \emph {et~al.}(2014)\citenamefont
		{Skrzypczyk}, \citenamefont {Short},\ and\ \citenamefont
		{Popescu}}]{skrzypczyk2014work}%
	\BibitemOpen
	\bibfield  {author} {\bibinfo {author} {\bibfnamefont {P.}~\bibnamefont
			{Skrzypczyk}}, \bibinfo {author} {\bibfnamefont {A.~J.}\ \bibnamefont
			{Short}}, \ and\ \bibinfo {author} {\bibfnamefont {S.}~\bibnamefont
			{Popescu}},\ }\bibinfo {title} {Work extraction and thermodynamics for
		individual quantum systems},\ \href@noop {} {\bibfield  {journal} {\bibinfo
			{journal} {Nature communications}\ }\textbf {\bibinfo {volume} {5}},\
		\bibinfo {pages} {4185} (\bibinfo {year} {2014})}\BibitemShut {NoStop}%
	\bibitem [{\citenamefont {Mazzoncini}\ \emph {et~al.}(2023)\citenamefont
		{Mazzoncini}, \citenamefont {Cavina}, \citenamefont {Andolina}, \citenamefont
		{Erdman},\ and\ \citenamefont {Giovannetti}}]{PhysRevA.107.032218}%
	\BibitemOpen
	\bibfield  {author} {\bibinfo {author} {\bibfnamefont {F.}~\bibnamefont
			{Mazzoncini}}, \bibinfo {author} {\bibfnamefont {V.}~\bibnamefont {Cavina}},
		\bibinfo {author} {\bibfnamefont {G.~M.}\ \bibnamefont {Andolina}}, \bibinfo
		{author} {\bibfnamefont {P.~A.}\ \bibnamefont {Erdman}}, \ and\ \bibinfo
		{author} {\bibfnamefont {V.}~\bibnamefont {Giovannetti}},\ }\bibinfo {title}
	{Optimal control methods for quantum batteries},\ \href {\doibase
		10.1103/PhysRevA.107.032218} {\bibfield  {journal} {\bibinfo  {journal}
			{Phys. Rev. A}\ }\textbf {\bibinfo {volume} {107}},\ \bibinfo {pages}
		{032218} (\bibinfo {year} {2023})}\BibitemShut {NoStop}%
	\bibitem [{\citenamefont {Yao}\ and\ \citenamefont
		{Shao}(2022)}]{PhysRevE.106.014138}%
	\BibitemOpen
	\bibfield  {author} {\bibinfo {author} {\bibfnamefont {Y.}~\bibnamefont
			{Yao}}\ and\ \bibinfo {author} {\bibfnamefont {X.~Q.}\ \bibnamefont {Shao}},\
	}\bibinfo {title} {Optimal charging of open spin-chain quantum batteries via
		homodyne-based feedback control},\ \href {\doibase
		10.1103/PhysRevE.106.014138} {\bibfield  {journal} {\bibinfo  {journal}
			{Phys. Rev. E}\ }\textbf {\bibinfo {volume} {106}},\ \bibinfo {pages}
		{014138} (\bibinfo {year} {2022})}\BibitemShut {NoStop}%
	\bibitem [{\citenamefont {Evangelakos}\ \emph {et~al.}(2024)\citenamefont
		{Evangelakos}, \citenamefont {Paspalakis},\ and\ \citenamefont
		{Stefanatos}}]{PhysRevA.110.052601}%
	\BibitemOpen
	\bibfield  {author} {\bibinfo {author} {\bibfnamefont {V.}~\bibnamefont
			{Evangelakos}}, \bibinfo {author} {\bibfnamefont {E.}~\bibnamefont
			{Paspalakis}}, \ and\ \bibinfo {author} {\bibfnamefont {D.}~\bibnamefont
			{Stefanatos}},\ }\bibinfo {title} {Fast charging of an Ising-spin-pair
		quantum battery using optimal control},\ \href {\doibase
		10.1103/PhysRevA.110.052601} {\bibfield  {journal} {\bibinfo  {journal}
			{Phys. Rev. A}\ }\textbf {\bibinfo {volume} {110}},\ \bibinfo {pages}
		{052601} (\bibinfo {year} {2024})}\BibitemShut {NoStop}%
	\bibitem [{\citenamefont {Ali}\ \emph {et~al.}(2024)\citenamefont {Ali},
		\citenamefont {Al-Kuwari}, \citenamefont {Hussain}, \citenamefont {Byrnes},
		\citenamefont {Rahim}, \citenamefont {Quach}, \citenamefont {Ghominejad},\
		and\ \citenamefont {Haddadi}}]{PhysRevA.110.052404}%
	\BibitemOpen
	\bibfield  {author} {\bibinfo {author} {\bibfnamefont {A.}~\bibnamefont
			{Ali}}, \bibinfo {author} {\bibfnamefont {S.}~\bibnamefont {Al-Kuwari}},
		\bibinfo {author} {\bibfnamefont {M.~I.}\ \bibnamefont {Hussain}}, \bibinfo
		{author} {\bibfnamefont {T.}~\bibnamefont {Byrnes}}, \bibinfo {author}
		{\bibfnamefont {M.~T.}\ \bibnamefont {Rahim}}, \bibinfo {author}
		{\bibfnamefont {J.~Q.}\ \bibnamefont {Quach}}, \bibinfo {author}
		{\bibfnamefont {M.}~\bibnamefont {Ghominejad}}, \ and\ \bibinfo {author}
		{\bibfnamefont {S.}~\bibnamefont {Haddadi}},\ }\bibinfo {title} {Ergotropy
		and capacity optimization in Heisenberg spin-chain quantum batteries},\ \href
	{\doibase 10.1103/PhysRevA.110.052404} {\bibfield  {journal} {\bibinfo
			{journal} {Phys. Rev. A}\ }\textbf {\bibinfo {volume} {110}},\ \bibinfo
		{pages} {052404} (\bibinfo {year} {2024})}\BibitemShut {NoStop}%
	\bibitem [{\citenamefont {Yao}\ and\ \citenamefont {Shao}(2025)}]{vqnk-kzqg}%
	\BibitemOpen
	\bibfield  {author} {\bibinfo {author} {\bibfnamefont {Y.}~\bibnamefont
			{Yao}}\ and\ \bibinfo {author} {\bibfnamefont {X.~Q.}\ \bibnamefont {Shao}},\
	}\bibinfo {title} {Reservoir-assisted quantum battery charging at finite
		temperatures},\ \href {\doibase 10.1103/vqnk-kzqg} {\bibfield  {journal}
		{\bibinfo  {journal} {Phys. Rev. A}\ }\textbf {\bibinfo {volume} {111}},\
		\bibinfo {pages} {062616} (\bibinfo {year} {2025})}\BibitemShut {NoStop}%
	\bibitem [{\citenamefont {Santos}(2021)}]{PhysRevE.103.042118}%
	\BibitemOpen
	\bibfield  {author} {\bibinfo {author} {\bibfnamefont {A.~C.}\ \bibnamefont
			{Santos}},\ }\bibinfo {title} {Quantum advantage of two-level batteries in
		the self-discharging process},\ \href {\doibase 10.1103/PhysRevE.103.042118}
	{\bibfield  {journal} {\bibinfo  {journal} {Phys. Rev. E}\ }\textbf {\bibinfo
			{volume} {103}},\ \bibinfo {pages} {042118} (\bibinfo {year}
		{2021})}\BibitemShut {NoStop}%
	\bibitem [{\citenamefont {Arjmandi}\ \emph
		{et~al.}(2022{\natexlab{b}})\citenamefont {Arjmandi}, \citenamefont
		{Mohammadi},\ and\ \citenamefont {Santos}}]{PhysRevE.105.054115}%
	\BibitemOpen
	\bibfield  {author} {\bibinfo {author} {\bibfnamefont {M.~B.}\ \bibnamefont
			{Arjmandi}}, \bibinfo {author} {\bibfnamefont {H.}~\bibnamefont {Mohammadi}},
		\ and\ \bibinfo {author} {\bibfnamefont {A.~C.}\ \bibnamefont {Santos}},\
	}\bibinfo {title} {Enhancing self-discharging process with disordered quantum
		batteries},\ \href {\doibase 10.1103/PhysRevE.105.054115} {\bibfield
		{journal} {\bibinfo  {journal} {Phys. Rev. E}\ }\textbf {\bibinfo {volume}
			{105}},\ \bibinfo {pages} {054115} (\bibinfo {year}
		{2022}{\natexlab{b}})}\BibitemShut {NoStop}%
	\bibitem [{\citenamefont {Xu}\ \emph {et~al.}(2024)\citenamefont {Xu},
		\citenamefont {Li}, \citenamefont {Zhu},\ and\ \citenamefont
		{Liu}}]{PhysRevE.109.054132}%
	\BibitemOpen
	\bibfield  {author} {\bibinfo {author} {\bibfnamefont {K.}~\bibnamefont
			{Xu}}, \bibinfo {author} {\bibfnamefont {H.-G.}\ \bibnamefont {Li}}, \bibinfo
		{author} {\bibfnamefont {H.-J.}\ \bibnamefont {Zhu}}, \ and\ \bibinfo
		{author} {\bibfnamefont {W.-M.}\ \bibnamefont {Liu}},\ }\bibinfo {title}
	{Inhibiting the self-discharging process of quantum batteries in
		non-Markovian noises},\ \href {\doibase 10.1103/PhysRevE.109.054132}
	{\bibfield  {journal} {\bibinfo  {journal} {Phys. Rev. E}\ }\textbf {\bibinfo
			{volume} {109}},\ \bibinfo {pages} {054132} (\bibinfo {year}
		{2024})}\BibitemShut {NoStop}%
	\bibitem [{\citenamefont {Xu}\ \emph {et~al.}(2021)\citenamefont {Xu},
		\citenamefont {Zhu}, \citenamefont {Zhang},\ and\ \citenamefont
		{Liu}}]{PhysRevE.104.064143}%
	\BibitemOpen
	\bibfield  {author} {\bibinfo {author} {\bibfnamefont {K.}~\bibnamefont
			{Xu}}, \bibinfo {author} {\bibfnamefont {H.-J.}\ \bibnamefont {Zhu}},
		\bibinfo {author} {\bibfnamefont {G.-F.}\ \bibnamefont {Zhang}}, \ and\
		\bibinfo {author} {\bibfnamefont {W.-M.}\ \bibnamefont {Liu}},\ }\bibinfo
	{title} {Enhancing the performance of an open quantum battery via environment
		engineering},\ \href {\doibase 10.1103/PhysRevE.104.064143} {\bibfield
		{journal} {\bibinfo  {journal} {Phys. Rev. E}\ }\textbf {\bibinfo {volume}
			{104}},\ \bibinfo {pages} {064143} (\bibinfo {year} {2021})}\BibitemShut
	{NoStop}%
	\bibitem [{\citenamefont {Kamin}\ \emph
		{et~al.}(2020{\natexlab{b}})\citenamefont {Kamin}, \citenamefont {Tabesh},
		\citenamefont {Salimi}, \citenamefont {Kheirandish},\ and\ \citenamefont
		{Santos}}]{Kamin_2020}%
	\BibitemOpen
	\bibfield  {author} {\bibinfo {author} {\bibfnamefont {F.~H.}\ \bibnamefont
			{Kamin}}, \bibinfo {author} {\bibfnamefont {F.~T.}\ \bibnamefont {Tabesh}},
		\bibinfo {author} {\bibfnamefont {S.}~\bibnamefont {Salimi}}, \bibinfo
		{author} {\bibfnamefont {F.}~\bibnamefont {Kheirandish}}, \ and\ \bibinfo
		{author} {\bibfnamefont {A.~C.}\ \bibnamefont {Santos}},\ }\bibinfo {title}
	{Non-Markovian effects on charging and self-discharging process of quantum
		batteries},\ \href {\doibase 10.1088/1367-2630/ab9ee2} {\bibfield  {journal}
		{\bibinfo  {journal} {New Journal of Physics}\ }\textbf {\bibinfo {volume}
			{22}},\ \bibinfo {pages} {083007} (\bibinfo {year}
		{2020}{\natexlab{b}})}\BibitemShut {NoStop}%
	\bibitem [{\citenamefont {Li}\ \emph {et~al.}(2022)\citenamefont {Li},
		\citenamefont {Shen},\ and\ \citenamefont {Yi}}]{Li:22}%
	\BibitemOpen
	\bibfield  {author} {\bibinfo {author} {\bibfnamefont {J.~L.}\ \bibnamefont
			{Li}}, \bibinfo {author} {\bibfnamefont {H.~Z.}\ \bibnamefont {Shen}}, \ and\
		\bibinfo {author} {\bibfnamefont {X.~X.}\ \bibnamefont {Yi}},\ }\bibinfo
	{title} {Quantum batteries in non-Markovian reservoirs},\ \href {\doibase
		10.1364/OL.471820} {\bibfield  {journal} {\bibinfo  {journal} {Opt. Lett.}\
		}\textbf {\bibinfo {volume} {47}},\ \bibinfo {pages} {5614} (\bibinfo {year}
		{2022})}\BibitemShut {NoStop}%
	\bibitem [{\citenamefont {Bhanja}\ \emph {et~al.}(2024)\citenamefont {Bhanja},
		\citenamefont {Tiwari},\ and\ \citenamefont
		{Banerjee}}]{PhysRevA.109.012224}%
	\BibitemOpen
	\bibfield  {author} {\bibinfo {author} {\bibfnamefont {G.}~\bibnamefont
			{Bhanja}}, \bibinfo {author} {\bibfnamefont {D.}~\bibnamefont {Tiwari}}, \
		and\ \bibinfo {author} {\bibfnamefont {S.}~\bibnamefont {Banerjee}},\
	}\bibinfo {title} {Impact of non-Markovian quantum Brownian motion on quantum
		batteries},\ \href {\doibase 10.1103/PhysRevA.109.012224} {\bibfield
		{journal} {\bibinfo  {journal} {Phys. Rev. A}\ }\textbf {\bibinfo {volume}
			{109}},\ \bibinfo {pages} {012224} (\bibinfo {year} {2024})}\BibitemShut
	{NoStop}%
	\bibitem [{\citenamefont {Bai}\ and\ \citenamefont
		{An}(2020)}]{PhysRevA.102.060201}%
	\BibitemOpen
	\bibfield  {author} {\bibinfo {author} {\bibfnamefont {S.-Y.}\ \bibnamefont
			{Bai}}\ and\ \bibinfo {author} {\bibfnamefont {J.-H.}\ \bibnamefont {An}},\
	}\bibinfo {title} {Floquet engineering to reactivate a dissipative quantum
		battery},\ \href {\doibase 10.1103/PhysRevA.102.060201} {\bibfield  {journal}
		{\bibinfo  {journal} {Phys. Rev. A}\ }\textbf {\bibinfo {volume} {102}},\
		\bibinfo {pages} {060201(R)} (\bibinfo {year} {2020})}\BibitemShut {NoStop}%
	\bibitem [{\citenamefont {Gonz\'alez-Tudela}\ and\ \citenamefont
		{Cirac}(2017)}]{PhysRevA.96.043811}%
	\BibitemOpen
	\bibfield  {author} {\bibinfo {author} {\bibfnamefont {A.}~\bibnamefont
			{Gonz\'alez-Tudela}}\ and\ \bibinfo {author} {\bibfnamefont {J.~I.}\
			\bibnamefont {Cirac}},\ }\bibinfo {title} {Markovian and non-Markovian
		dynamics of quantum emitters coupled to two-dimensional structured
		reservoirs},\ \href {\doibase 10.1103/PhysRevA.96.043811} {\bibfield
		{journal} {\bibinfo  {journal} {Phys. Rev. A}\ }\textbf {\bibinfo {volume}
			{96}},\ \bibinfo {pages} {043811} (\bibinfo {year} {2017})}\BibitemShut
	{NoStop}%
	\bibitem [{\citenamefont {Gamel}\ and\ \citenamefont
		{James}(2010)}]{PhysRevA.82.052106}%
	\BibitemOpen
	\bibfield  {author} {\bibinfo {author} {\bibfnamefont {O.}~\bibnamefont
			{Gamel}}\ and\ \bibinfo {author} {\bibfnamefont {D.~F.~V.}\ \bibnamefont
			{James}},\ }\bibinfo {title} {Time-averaged quantum dynamics and the validity
		of the effective Hamiltonian model},\ \href {\doibase
		10.1103/PhysRevA.82.052106} {\bibfield  {journal} {\bibinfo  {journal} {Phys.
				Rev. A}\ }\textbf {\bibinfo {volume} {82}},\ \bibinfo {pages} {052106}
		(\bibinfo {year} {2010})}\BibitemShut {NoStop}%
	\bibitem [{\citenamefont {Shao}\ \emph {et~al.}(2017)\citenamefont {Shao},
		\citenamefont {Wu},\ and\ \citenamefont {Feng}}]{PhysRevA.95.032124}%
	\BibitemOpen
	\bibfield  {author} {\bibinfo {author} {\bibfnamefont {W.}~\bibnamefont
			{Shao}}, \bibinfo {author} {\bibfnamefont {C.}~\bibnamefont {Wu}}, \ and\
		\bibinfo {author} {\bibfnamefont {X.-L.}\ \bibnamefont {Feng}},\ }\bibinfo
	{title} {Generalized James' effective Hamiltonian method},\ \href {\doibase
		10.1103/PhysRevA.95.032124} {\bibfield  {journal} {\bibinfo  {journal} {Phys.
				Rev. A}\ }\textbf {\bibinfo {volume} {95}},\ \bibinfo {pages} {032124}
		(\bibinfo {year} {2017})}\BibitemShut {NoStop}%
	\bibitem [{sup()}]{supp}%
	\BibitemOpen
	\bibinfo {title} {See supplement},\ \href@noop {} {\ }\BibitemShut {NoStop}%
	\bibitem [{\citenamefont {Popkov}\ and\ \citenamefont
		{Presilla}(2021)}]{PhysRevLett.126.190402}%
	\BibitemOpen
	\bibfield  {author} {\bibinfo {author} {\bibfnamefont {V.}~\bibnamefont
			{Popkov}}\ and\ \bibinfo {author} {\bibfnamefont {C.}~\bibnamefont
			{Presilla}},\ }\bibinfo {title} {Full Spectrum of the Liouvillian of Open
		Dissipative Quantum Systems in the Zeno Limit},\ \href {\doibase
		10.1103/PhysRevLett.126.190402} {\bibfield  {journal} {\bibinfo  {journal}
			{Phys. Rev. Lett.}\ }\textbf {\bibinfo {volume} {126}},\ \bibinfo {pages}
		{190402} (\bibinfo {year} {2021})}\BibitemShut {NoStop}%
\end{thebibliography}


\end{document}